\input epsf
\input amssym
\input\jobname.intref

\newfam\scrfam
\batchmode\font\tenscr=rsfs10 \errorstopmode
\ifx\tenscr\nullfont
        \message{rsfs script font not available. Replacing with calligraphic.}
        \def\scr{\cal}
\else   
        \font\sevenscr=rsfs7
        \font\fivescr=rsfs5
        \skewchar\tenscr='177 \skewchar\sevenscr='177 \skewchar\fivescr='177
        \textfont\scrfam=\tenscr \scriptfont\scrfam=\sevenscr
        \scriptscriptfont\scrfam=\fivescr
        \def\scr{\fam\scrfam}
        \def\cal{\scr}
\fi
\catcode`\@=11
\newfam\frakfam
\batchmode\font\tenfrak=eufm10 \errorstopmode
\ifx\tenfrak\nullfont
        \message{eufm font not available. Replacing with italic.}
        \def\frak{\it}
\else
	
	\font\sevenfrak=eufm7 \font\fivefrak=eufm5
        
	\textfont\frakfam=\tenfrak
	\scriptfont\frakfam=\sevenfrak \scriptscriptfont\frakfam=\fivefrak
	\def\frak{\fam\frakfam}
\fi
\catcode`\@=\active
\newfam\msbfam
\batchmode\font\twelvemsb=msbm10 scaled\magstep1 \errorstopmode
\ifx\twelvemsb\nullfont\def\Bbb{\bf}
        
	\font\eightbbb=cmb10 at 8pt
	\message{Blackboard bold not available. Replacing with boldface.}
\else   \catcode`\@=11
        \font\tenmsb=msbm10 \font\sevenmsb=msbm7 \font\fivemsb=msbm5
        \textfont\msbfam=\tenmsb
        \scriptfont\msbfam=\sevenmsb \scriptscriptfont\msbfam=\fivemsb
        \def\Bbb{\relax\expandafter\Bbb@}
        \def\Bbb@#1{{\Bbb@@{#1}}}
        \def\Bbb@@#1{\fam\msbfam\relax#1}
        \catcode`\@=\active
	
	\font\eightbbb=msbm8
\fi
        \font\fivemi=cmmi5
        \font\sixmi=cmmi6
        \font\eightrm=cmr8              \def\xrm{\eightrm}
        \font\eightbf=cmbx8             \def\xbf{\eightbf}
        \font\eightit=cmti10 at 8pt     \def\xit{\eightit}

        \font\eighttt=cmtt8

        \font\eightcp=cmcsc8    
                      \def\xold{\eighti}
        \font\eightmi=cmmi8
                     \def\xbold{\eightib}
        \font\teni=cmmi10               \def\old{\teni}
        \font\tencp=cmcsc10

        \font\twelvecp=cmcsc10 scaled\magstep1
        
        \font\sixrm=cmr6
        \font\fiverm=cmr5

        \font\eightsy=cmsy8
        \font\sixsy=cmsy6
        \font\eightsl=cmsl8

        \font\sixbf=cmbx6

	 at10pt	
	\font\twelvehelvbold=phvb at12pt
	 at14pt
	\font\sixteenhelvbold=phvb at16pt
	 at16pt



\def\xbold{\xbf}
\def\xold{\xrm}


\def\noblackbox{\overfullrule=0pt}
\noblackbox

\def\eightpoint{
\def\rm{\fam0\eightrm}
\textfont0=\eightrm \scriptfont0=\sixrm \scriptscriptfont0=\fiverm
\textfont1=\eightmi  \scriptfont1=\sixmi  \scriptscriptfont1=\fivemi
\textfont2=\eightsy \scriptfont2=\sixsy \scriptscriptfont2=\fivesy
\textfont3=\tenex   \scriptfont3=\tenex \scriptscriptfont3=\tenex
\textfont\itfam=\eightit \def\it{\fam\itfam\eightit}
\textfont\slfam=\eightsl \def\sl{\fam\slfam\eightsl}
\textfont\ttfam=\eighttt \def\tt{\fam\ttfam\eighttt}
\textfont\bffam=\eightbf \scriptfont\bffam=\sixbf 
                         \scriptscriptfont\bffam=\fivebf
                         \def\bf{\fam\bffam\eightbf}
\normalbaselineskip=10pt}



\newtoks\headtext
\headline={\ifnum\pageno=1\hfill\else
	\ifodd\pageno
        \noindent{\eightcp\the\headtext}{ }\dotfill{ }{\old\folio}
	\else\noindent{\old\folio}{ }\dotfill{ }{\eightcp\the\headtext}\fi
	\fi}
\def\makeheadline{\vbox to 0pt{\vss\noindent\the\headline\break
\hbox to\hsize{\hfill}}
        \vskip2\baselineskip}
\newcount\infootnote
\infootnote=0
\newcount\footnotecount
\footnotecount=1
\def\foot#1{\infootnote=1
\footnote{${}^{\the\footnotecount}$}{\vtop{\baselineskip=.75\baselineskip
\advance\hsize by
-\parindent{\eightpoint\rm\hskip-\parindent
#1}\hfill\vskip\parskip}}\infootnote=0\global\advance\footnotecount by
1\hskip2pt}
\newcount\refcount
\refcount=1
\newwrite\refwrite
\def\oldsize{\ifnum\infootnote=1\xold\else\old\fi}
\def\ref#1#2{
	\def#1{{{\oldsize\the\refcount}}\ifnum\the\refcount=1\immediate\openout\refwrite=\jobname.refs\fi\immediate\write\refwrite{\item{[{\xold\the\refcount}]} 
	#2\hfill\par\vskip-2pt}\xdef#1{{\noexpand\oldsize\the\refcount}}\global\advance\refcount by 1}
	}
\def\refout{\eightpoint\catcode`\@=11
        \xrm\immediate\closeout\refwrite
        \vskip2\baselineskip
        {\noindent\twelvecp References}\hfill\vskip\baselineskip
        \baselineskip=.75\baselineskip
        \input\jobname.refs
        \baselineskip=4\baselineskip \divide\baselineskip by 3
        \catcode`\@=\active\rm}

\def\skipref#1{\hbox to15pt{\phantom{#1}\hfill}\hskip-15pt}

\def\hepth#1{\href{http://xxx.lanl.gov/abs/hep-th/#1}{arXiv:\allowbreak
hep-th\slash{\xold#1}}}

\def\arxiv#1#2{\href{http://arxiv.org/abs/#1.#2}{arXiv:\allowbreak
{\xold#1}.{\xold#2}}} 
 
\def\jhep#1#2#3#4{\href{http://jhep.sissa.it/stdsearch?paper=#2\%28#3\%29#4}{J. High Energy Phys. {\xbold #1#2} ({\xold#3}) {\xold#4}}}

\def\FP#1#2#3{Fortsch. Phys. {\xbold#1} ({\xold#2}) {\xold#3}}

\def\JMP#1#2#3{J. Math. Phys. {\xbold#1} ({\xold#2}) {\xold#3}}
\def\JPA#1#2#3{J. Phys. {\xbf A}{\xbold#1} ({\xold#2}) {\xold#3}}

\def\NPB#1#2#3{Nucl. Phys. {\xbf B}{\xbold#1} ({\xold#2}) {\xold#3}}

\def\PLB#1#2#3{Phys. Lett. {\xbf B}{\xbold#1} ({\xold#2}) {\xold#3}}

\def\PRD#1#2#3{Phys. Rev. {\xbf D}{\xbold#1} ({\xold#2}) {\xold#3}}
\def\PRL#1#2#3{Phys. Rev. Lett. {\xbold#1} ({\xold#2}) {\xold#3}}

\newcount\sectioncount
\sectioncount=0
\def\section#1#2{\global\eqcount=0
	\global\subsectioncount=0
        \global\advance\sectioncount by 1
	\ifnum\sectioncount>1
	        \vskip2\baselineskip
	\fi
\noindent{\twelvecp\the\sectioncount. #2}\par\nobreak
       \vskip.5\baselineskip\noindent
        \xdef#1{{\old\the\sectioncount}}}
\newcount\subsectioncount
\def\subsection#1#2{\global\advance\subsectioncount by 1
\vskip.75\baselineskip\noindent\line{\tencp\the\sectioncount.\the\subsectioncount. #2\hfill}\nobreak\vskip.4\baselineskip\nobreak\noindent\xdef#1{{\old\the\sectioncount}.{\old\the\subsectioncount}}}
\def\immediatesubsection#1#2{\global\advance\subsectioncount by 1
\vskip-\baselineskip\noindent
\line{\tencp\the\sectioncount.\the\subsectioncount. #2\hfill}
	\vskip.5\baselineskip\noindent
	\xdef#1{{\old\the\sectioncount}.{\old\the\subsectioncount}}}
\newcount\subsubsectioncount
\def\subsubsection#1#2{\global\advance\subsubsectioncount by 1
\vskip.75\baselineskip\noindent\line{\tencp\the\sectioncount.\the\subsectioncount.\the\subsubsectioncount. #2\hfill}\nobreak\vskip.4\baselineskip\nobreak\noindent\xdef#1{{\old\the\sectioncount}.{\old\the\subsectioncount}.{\old\the\subsubsectioncount}}}
\newcount\appendixcount
\appendixcount=0
\def\appendix#1{\global\eqcount=0
        \global\advance\appendixcount by 1
        \vskip2\baselineskip\noindent
        \ifnum\the\appendixcount=1
        {\twelvecp Appendix A: #1}\par\nobreak
                        \vskip.5\baselineskip\noindent\fi
        \ifnum\the\appendixcount=2
        {\twelvecp Appendix B: #1}\par\nobreak
                        \vskip.5\baselineskip\noindent\fi
        \ifnum\the\appendixcount=3
        {\twelvecp Appendix C: #1}\par\nobreak
                        \vskip.5\baselineskip\noindent\fi}
\def\acknowledgements{\immediate\write\contentswrite{\item{}\hbox
        to\contentlength{Acknowledgements\dotfill\the\pageno}}
        \vskip2\baselineskip\noindent
        \underbar{\it Acknowledgements:}\ }
\newcount\eqcount
\eqcount=0
\def\Eqn#1{\global\advance\eqcount by 1
\ifnum\the\sectioncount=0
	\xdef#1{{\noexpand\oldsize\the\eqcount}}
	\eqno({\oldstyle\the\eqcount})
\else
        \ifnum\the\appendixcount=0
\xdef#1{{\noexpand\oldsize\the\sectioncount}.{\noexpand\oldsize\the\eqcount}}
                \eqno({\oldstyle\the\sectioncount}.{\oldstyle\the\eqcount})\fi
        \ifnum\the\appendixcount=1
	        \xdef#1{{\noexpand\oldstyle A}.{\noexpand\oldstyle\the\eqcount}}
                \eqno({\oldstyle A}.{\oldstyle\the\eqcount})\fi
        \ifnum\the\appendixcount=2
	        \xdef#1{{\noexpand\oldstyle B}.{\noexpand\oldstyle\the\eqcount}}
                \eqno({\oldstyle B}.{\oldstyle\the\eqcount})\fi
        \ifnum\the\appendixcount=3
	        \xdef#1{{\noexpand\oldstyle C}.{\noexpand\oldstyle\the\eqcount}}
                \eqno({\oldstyle C}.{\oldstyle\the\eqcount})\fi
\fi}
\def\eqn{\global\advance\eqcount by 1
\ifnum\the\sectioncount=0
	\eqno({\oldstyle\the\eqcount})
\else
        \ifnum\the\appendixcount=0
                \eqno({\oldstyle\the\sectioncount}.{\oldstyle\the\eqcount})\fi
        \ifnum\the\appendixcount=1
                \eqno({\oldstyle A}.{\oldstyle\the\eqcount})\fi
        \ifnum\the\appendixcount=2
                \eqno({\oldstyle B}.{\oldstyle\the\eqcount})\fi
        \ifnum\the\appendixcount=3
                \eqno({\oldstyle C}.{\oldstyle\the\eqcount})\fi
\fi}
\def\multi{\global\advance\eqcount by 1}
\def\multieqn#1{({\oldstyle\the\sectioncount}.{\oldstyle\the\eqcount}\hbox{#1})}
\def\multiEqn#1#2{\xdef#1{{\oldstyle\the\sectioncount}.{\old\the\eqcount}#2}
        ({\oldstyle\the\sectioncount}.{\oldstyle\the\eqcount}\hbox{#2})}
\def\multiEqnAll#1{\xdef#1{{\oldstyle\the\sectioncount}.{\old\the\eqcount}}}
\newcount\tablecount
\tablecount=0
\def\Table#1#2#3{\global\advance\tablecount by 1
\immediate\write\intrefwrite{\def\noexpand#1{{\noexpand\oldsize\the\tablecount}}}
       \vtop{\vskip2\parskip
       \centerline{#2}
       \vskip5\parskip
       \centerline{\it Table \the\tablecount: #3}
       \vskip2\parskip}}
\newcount\figurecount
\figurecount=0
\def\Figure#1#2#3{\global\advance\figurecount by 1
\immediate\write\intrefwrite{\def\noexpand#1{{\noexpand\oldsize\the\figurecount}}}
       \vtop{\vskip2\parskip
       \centerline{#2}
       \vskip4\parskip
       \centerline{\it Figure \the\figurecount: #3}
       \vskip3\parskip}}
\def\TextFigure#1#2#3{\global\advance\figurecount by 1
\immediate\write\intrefwrite{\def\noexpand#1{{\noexpand\oldsize\the\figurecount}}}
       \vtop{\vskip2\parskip
       \centerline{#2}
       \vskip4\parskip
       {\narrower\noindent\it Figure \the\figurecount: #3\smallskip}
       \vskip3\parskip}}
\newtoks\url
\def\Href#1#2{\catcode`\#=12\url={#1}\catcode`\#=\active#2}
\def\href#1#2{{#2}}

\parskip=3.5pt plus .3pt minus .3pt
\baselineskip=14pt plus .1pt minus .05pt
\lineskip=.5pt plus .05pt minus .05pt
\lineskiplimit=.5pt
\abovedisplayskip=18pt plus 4pt minus 2pt
\belowdisplayskip=\abovedisplayskip
\hsize=14cm
\vsize=19cm
\hoffset=1.5cm
\voffset=1.8cm
\frenchspacing
\footline={}
\raggedbottom

\newskip\origparindent
\origparindent=\parindent

\def\*{\partial}
\def\punkt{\;\,.}
\def\komma{\;\,,}

\def\={\!=\!}
\def\small#1{{\hbox{$#1$}}}

\def\fraction#1{\small{1\over#1}}
\def\fr{\fraction}
\def\Fraction#1#2{\small{#1\over#2}}
\def\Fr{\Fraction}

\def\eg{{\it e.g.}}

\def\ie{{\it i.e.}}

\def\bra#1{\langle#1|}
\def\ket#1{|#1\rangle}

\def\ketket#1{\ket{\ket{#1}\hskip-1pt}}

\def\II{I\hskip-.8pt I}

\def\RR{{\Bbb R}}


\def\appendix#1#2{\global\eqcount=0
        \global\advance\appendixcount by 1
        \vskip2\baselineskip\noindent
        \ifnum\the\appendixcount=1
        \immediate\write\intrefwrite{\def\noexpand#1{A}}
        {\twelvecp Appendix A: #2}\par\nobreak
                        \vskip.5\baselineskip\noindent\fi
        \ifnum\the\appendixcount=2
        {\twelvecp Appendix B: #2}\par\nobreak
                        \vskip.5\baselineskip\noindent\fi
        \ifnum\the\appendixcount=3
        {\twelvecp Appendix C: #2}\par\nobreak
                        \vskip.5\baselineskip\noindent\fi}

\def\textfrac#1#2{\raise .45ex\hbox{\the\scriptfont0 #1}\nobreak\hskip-1pt/\hskip-1pt\hbox{\the\scriptfont0 #2}}

\def\LL{{\cal L}}


\def\frac{\Fr}

\def\mathbb{\Bbb}



\def\LL{{\cal L}}

\def\LL{{\cal L}}

\def\LL{{\cal L}}

\def\fg{{\frak g}}




\catcode`@=11
\def\openupnormal{\afterassignment\@penupnormal\dimen@=}
\def\@penupnormal{\advance\normallineskip\dimen@
  \advance\normalbaselineskip\dimen@
  \advance\normallineskiplimit\dimen@}
\catcode`@=12

\def\EqMatrix{\let\quad\enspace\openupnormal6pt\matrix}



\def\textfrac#1#2{\raise .45ex\hbox{\the\scriptfont0 #1}\nobreak\hskip-1pt/\hskip-1pt\hbox{\the\scriptfont0 #2}}

\def\LL{{\cal L}}


\def\frac{\Fr}

\def\mathbb{\Bbb}

\newskip\scrskip
\scrskip=-.6pt plus 0pt minus .1pt


\newwrite\intrefwrite
\immediate\openout\intrefwrite=\jobname.intref

\newwrite\contentswrite

\newdimen\sublength
\sublength=\hsize 
\advance\sublength by -\parindent

\newdimen\contentlength
\contentlength=\sublength

\advance\sublength by -\parindent

\def\section#1#2{\global\eqcount=0
	\global\subsectioncount=0
        \global\advance\sectioncount by 1
\ifnum\the\sectioncount=1\immediate\openout\contentswrite=\jobname.contents\fi
\immediate\write\contentswrite{\item{\the\sectioncount.}\hbox to\contentlength{#2\dotfill\the\pageno}}
	\ifnum\sectioncount>1
	        \vskip2\baselineskip
	\fi
\immediate\write\intrefwrite{\def\noexpand#1{{\noexpand\oldsize\the\sectioncount}}}\noindent{\twelvecp\the\sectioncount. #2}\par\nobreak
       \vskip.5\baselineskip\noindent}

\def\subsection#1#2{\global\advance\subsectioncount by 1
\immediate\write\contentswrite{\itemitem{\the\sectioncount.\the\subsectioncount.}\hbox
to\sublength{#2\dotfill\the\pageno}}
\immediate\write\intrefwrite{\def\noexpand#1{{\noexpand\oldsize\the\sectioncount}.{\noexpand\oldsize\the\subsectioncount}}}\vskip.75\baselineskip\noindent\line{\tencp\the\sectioncount.\the\subsectioncount. #2\hfill}\nobreak\vskip.4\baselineskip\nobreak\noindent}

\def\immediatesubsection#1#2{\global\advance\subsectioncount by 1
\immediate\write\contentswrite{\itemitem{\the\sectioncount.\the\subsectioncount.}\hbox
to\sublength{#2\dotfill\the\pageno}}
\immediate\write\intrefwrite{\def\noexpand#1{{\noexpand\oldsize\the\sectioncount}.{\noexpand\oldsize\the\subsectioncount}}}
\vskip-\baselineskip\noindent
\line{\tencp\the\sectioncount.\the\subsectioncount. #2\hfill}
	\vskip.5\baselineskip\noindent}

\def\contentsout{\catcode`\@=11
        \vskip2\baselineskip
        {\noindent\twelvecp Contents}\hfill\vskip\baselineskip
        \input\jobname.contents
        \catcode`\@=\active\rm
\vskip3\baselineskip
}

\def\refout{\eightpoint\catcode`\@=11
        \immediate\write\contentswrite{\item{}\hbox to\contentlength{References\dotfill\the\pageno}}
        \xrm\immediate\closeout\refwrite
        \vskip2\baselineskip
        {\noindent\twelvecp References}\hfill\vskip\baselineskip
        \baselineskip=.75\baselineskip
        \input\jobname.refs
        \baselineskip=4\baselineskip \divide\baselineskip by 3
        \catcode`\@=\active\rm}


\ref\Tseytlin{A.A.~Tseytlin,
  {\xit ``Duality symmetric closed string theory and interacting
  chiral scalars''}, 
  \NPB{350}{1991}{395}.}

\ref\SiegelI{W.~Siegel,
  {\xit ``Two vierbein formalism for string inspired axionic gravity''},
  \PRD{47}{1993}{5453}
  [\hepth{9302036}].}

\ref\SiegelII{ W.~Siegel,
  {\xit ``Superspace duality in low-energy superstrings''},
  \PRD{48}{1993}{2826}
  [\hepth{9305073}].}

\ref\SiegelIII{W.~Siegel,
  {\xit ``Manifest duality in low-energy superstrings''},
  in Berkeley 1993, Proceedings, Strings '93 353
  [\hepth{9308133}].}

\ref\HullDoubled{C.M. Hull, {\xit ``Doubled geometry and
T-folds''}, \jhep{07}{07}{2007}{080}
[\hepth{0605149}].}

\ref\HullT{C.M. Hull, {\xit ``A geometry for non-geometric string
backgrounds''}, \jhep{05}{10}{2005}{065} [\hepth{0406102}].}

\ref\HullM{C.M. Hull, {\xit ``Generalised geometry for M-theory''},
\jhep{07}{07}{2007}{079} [\hepth{0701203}].}

\ref\HullZwiebachDFT{C. Hull and B. Zwiebach, {\xit ``Double field
theory''}, \jhep{09}{09}{2009}{99} [\arxiv{0904}{4664}].}

\ref\HohmHullZwiebachI{O. Hohm, C.M. Hull and B. Zwiebach, {\xit ``Background
independent action for double field
theory''}, \jhep{10}{07}{2010}{016} [\arxiv{1003}{5027}].}

\ref\HohmHullZwiebachII{O. Hohm, C.M. Hull and B. Zwiebach, {\xit
``Generalized metric formulation of double field theory''},
\jhep{10}{08}{2010}{008} [\arxiv{1006}{4823}].} 

\ref\HohmKwak{O. Hohm and S.K. Kwak, {\xit ``$N=1$ supersymmetric
double field theory''}, \jhep{12}{03}{2012}{080} [\arxiv{1111}{7293}].}

\ref\HohmKwakFrame{O. Hohm and S.K. Kwak, {\xit ``Frame-like geometry
of double field theory''}, \JPA{44}{2011}{085404} [\arxiv{1011}{4101}].}

\ref\JeonLeeParkI{I. Jeon, K. Lee and J.-H. Park, {\xit ``Differential
geometry with a projection: Application to double field theory''},
\jhep{11}{04}{2011}{014} [\arxiv{1011}{1324}].}

\ref\JeonLeeParkII{I. Jeon, K. Lee and J.-H. Park, {\xit ``Stringy
differential geometry, beyond Riemann''}, 
\PRD{84}{2011}{044022} [\arxiv{1105}{6294}].}

\ref\JeonLeeParkIII{I. Jeon, K. Lee and J.-H. Park, {\xit
``Supersymmetric double field theory: stringy reformulation of supergravity''},
\PRD{85}{2012}{081501} [\arxiv{1112}{0069}].}

\ref\HohmZwiebachLarge{O. Hohm and B. Zwiebach, {\xit ``Large gauge
transformations in double field theory''}, \jhep{13}{02}{2013}{075}
[\arxiv{1207}{4198}].} 

\ref\Park{J.-H.~Park,
  {\xit ``Comments on double field theory and diffeomorphisms''},
  \jhep{13}{06}{2013}{098}
  [\arxiv{1304}{5946}].}

\ref\BermanCederwallPerry{D.S. Berman, M. Cederwall and M.J. Perry,
{\xit ``Global aspects of double geometry''}, 
\jhep{14}{09}{2014}{66} [\arxiv{1401}{1311}].}

\ref\PachecoWaldram{P.P. Pacheco and D. Waldram, {\xit ``M-theory,
exceptional generalised geometry and superpotentials''},
\jhep{08}{09}{2008}{123} [\arxiv{0804}{1362}].}

\ref\Hillmann{C. Hillmann, {\xit ``Generalized $E_{7(7)}$ coset
dynamics and $D=11$ supergravity''}, \jhep{09}{03}{2009}{135}
[\arxiv{0901}{1581}].}

\ref\BermanPerryGen{D.S. Berman and M.J. Perry, {\xit ``Generalised
geometry and M-theory''}, \jhep{11}{06}{2011}{074} [\arxiv{1008}{1763}].}    

\ref\BermanGodazgarPerry{D.S. Berman, H. Godazgar and M.J. Perry,
{\xit ``SO(5,5) duality in M-theory and generalized geometry''},
\PLB{700}{2011}{65} [\arxiv{1103}{5733}].} 

\ref\BermanMusaevPerry{D.S. Berman, E.T. Musaev and M.J. Perry,
{\xit ``Boundary terms in generalized geometry and doubled field theory''},
\PLB{706}{2011}{228} [\arxiv{1110}{97}].} 

\ref\BermanGodazgarGodazgarPerry{D.S. Berman, H. Godazgar, M. Godazgar  
and M.J. Perry,
{\xit ``The local symmetries of M-theory and their formulation in
generalised geometry''}, \jhep{12}{01}{2012}{012}
[\arxiv{1110}{3930}].} 

\ref\BermanGodazgarPerryWest{D.S. Berman, H. Godazgar, M.J. Perry and
P. West,
{\xit ``Duality invariant actions and generalised geometry''}, 
\jhep{12}{02}{2012}{108} [\arxiv{1111}{0459}].} 

\ref\CoimbraStricklandWaldram{A. Coimbra, C. Strickland-Constable and
D. Waldram, {\xit ``$E_{d(d)}\times\hbox{\eightbbb R}^+$ generalised geometry,
connections and M theory'' }, \jhep{14}{02}{2014}{054} [\arxiv{1112}{3989}].} 

\ref\CoimbraStricklandWaldramII{A. Coimbra, C. Strickland-Constable and
D. Waldram, {\xit ``Supergravity as generalised geometry II:
$E_{d(d)}\times\hbox{\eightbbb R}^+$ and M theory''}, 
\jhep{14}{03}{2014}{019} [\arxiv{1212}{1586}].}  

\ref\JeonLeeParkSuh{I. Jeon, K. Lee, J.-H. Park and Y. Suh, {\xit
``Stringy unification of Type IIA and IIB supergravities under N=2
D=10 supersymmetric double field theory''}, \PLB{723}{2013}{245}
[\arxiv{1210}{5048}].} 

\ref\JeonLeeParkRR{I. Jeon, K. Lee and J.-H. Park, {\xit
``Ramond--Ramond cohomology and O(D,D) T-duality''},
\jhep{12}{09}{2012}{079} [\arxiv{1206}{3478}].} 

\ref\BermanCederwallKleinschmidtThompson{D.S. Berman, M. Cederwall,
A. Kleinschmidt and D.C. Thompson, {\xit ``The gauge structure of
generalised diffeomorphisms''}, \jhep{13}{01}{2013}{64} [\arxiv{1208}{5884}].}

\ref\ParkSuh{J.-H. Park and Y. Suh, {\xit ``U-geometry: SL(5)''},
\jhep{14}{06}{2014}{102} [\arxiv{1302}{1652}].} 

\ref\CederwallI{M.~Cederwall, J.~Edlund and A.~Karlsson,
  {\xit ``Exceptional geometry and tensor fields''},
  \jhep{13}{07}{2013}{028}
  [\arxiv{1302}{6736}].}

\ref\CederwallII{ M.~Cederwall,
  {\xit ``Non-gravitational exceptional supermultiplets''},
  \jhep{13}{07}{2013}{025}
  [\arxiv{1302}{6737}].}

\ref\HohmSamtlebenI{O.~Hohm and H.~Samtleben,
  {\xit ``Exceptional field theory I: $E_{6(6)}$ covariant form of
  M-theory and type IIB''}, 
  \PRD{89}{2014}{066016} [\arxiv{1312}{0614}].}

\ref\HohmSamtlebenII{O.~Hohm and H.~Samtleben,
  {\xit ``Exceptional field theory II: $E_{7(7)}$''},
  \PRD{89}{2014}{066016} [\arxiv{1312}{4542}].}

\ref\HohmSamtlebenIII{O. Hohm and H. Samtleben, {\xit ``Exceptional field
theory III: $E_{8(8)}$''}, \PRD{90}{2014}{066002} [\arxiv{1406}{3348}].}

\ref\KachruNew{S. Kachru, M.B. Schulz, P.K. Tripathy and S.P. Trivedi,
{\xit ``New supersymmetric string compactifications''}, 
\jhep{03}{03}{2003}{061} [\hepth{0211182}].}

\ref\Condeescu{C. Condeescu, I. Florakis, C. Kounnas and D. L\"ust, 
{\xit ``Gauged supergravities and non-geometric $Q$/$R$-fluxes from
asymmetric orbifold CFT's''}, 
\jhep{13}{10}{2013}{057} [\arxiv{1307}{0999}].}


\ref\HasslerLust{F. Hassler and D. L\"ust, {\xit ``Consistent
compactification of double field theory on non-geometric flux
backgrounds''}, \jhep{14}{05}{2014}{085} [\arxiv{1401}{5068}].}

\ref\CederwallGeometryBehind{M. Cederwall, {\xit ``The geometry behind
double geometry''}, 
\jhep{14}{09}{2014}{70} [\arxiv{1402}{2513}].}

\ref\CederwallDuality{M. Cederwall, {\xit ``T-duality and
non-geometric solutions from double geometry''}, \FP{62}{2014}{942}
[\arxiv{1409}{4463}].} 

\ref\CederwallRosabal{M. Cederwall and J.A. Rosabal, ``$E_8$
geometry'', \jhep{15}{07}{2015}{007}, [\arxiv{1504}{04843}].}

\ref\HohmKwakZwiebachI{O. Hohm, S.K. Kwak and B. Zwiebach, {\xit
``Unification of type II strings and T-duality''}, \PRL{107}{2011}{171603}
[\arxiv{1106}{5452}].}  

\ref\HohmKwakZwiebachII{O. Hohm, S.K. Kwak and B. Zwiebach, {\xit
``Double field theory of type II strings''}, \jhep{11}{09}{2011}{013}
[\arxiv{1107}{0008}].}  

\ref\HohmZwiebachGeometry{O. Hohm and B. Zwiebach, {\xit ``Towards an
invariant geometry of double field theory''}, \arxiv{1212}{1736}.} 

\ref\CartanSpinors{E. Cartan, {\xit ``Le\hskip.5pt,\hskip-3.5pt cons sur
la th\'eorie des spineurs''} (Hermann, Paris, 1937).}

\ref\KacBook{V. G. Kac, {\xit ``Infinite-dimensional Lie
algebras''}, Cambridge Univ. Press
({\xold1990}).}

\ref\KacSuperalgebras{V.G. Kac, {\xit ``Classification of simple Lie
superalgebras''}, Funktsional. Anal. i Prilozhen. {\xbold9}
({\xold1975}) {\xold91}.}

\ref\CederwallExceptionalTwistors{M. Cederwall, {\xit ``Twistors and
supertwistors for exceptional field
theory''}, \jhep{15}{12}{2015}{123} [\arxiv{1510}{02298}].}

\ref\CederwallDoubleSuperGeometry{M. Cederwall, {\xit ``Double
supergeometry''}, \jhep{16}{06}{2016}{155} [\arxiv{1603}{04684}].}

\ref\CederwallPalmkvistBorcherds{M. Cederwall and J. Palmkvist, {\xit
``Superalgebras, constraints and partition functions''},
\jhep{08}{15}{2015}{36} [\arxiv{1503}{06215}].}

\ref\PalmkvistBorcherds{J. Palmkvist, {\xit
``Exceptional geometry and Borcherds superalgebras''},
\jhep{15}{11}{2015}{032} [\arxiv{1507}{08828}].}

\ref\StricklandConstable{C. Strickland-Constable,
  {\xit ``Subsectors, Dynkin diagrams and new generalised geometries''},
  \jhep{17}{08}{2017}{144}
  [\arxiv{1310}{4196}].}

\ref\Baraglia{D. Baraglia. {\xit ``Leibniz algebroids, twistings and
  exceptional generalized geometry''}, J. Geom. 
         Phys. {\xbf62} (2012) 903 [\arxiv{1101}{0856}].}

\ref\ENinePaper{G. Bossard, M. Cederwall, A. Kleinschmidt,
J. Palmkvist and H. Samtleben, {\xit ``Generalised diffeomorphisms for
$E_9$''}, \arxiv{1708}{08936}.}

\ref\HohmSamtlebenIII{O. Hohm and H. Samtleben, {\xit ``Exceptional field
theory III: $E_{8(8)}$''}, \PRD{90}{2014}{066002} [\arxiv{1406}{3348}].}

\ref\CederwallRosabal{M. Cederwall and J.A. Rosabal, ``$E_8$
geometry'', \jhep{15}{07}{2015}{007}, [\arxiv{1504}{04843}].}

\ref\PetersonKac{D.H. Peterson and V.G. Kac, {\xit ``Infinite flag
varieties and conjugacy theorems''}, Proc. Natl. Acad. Sci. {\xbf80}
(1983) 1778.}

\ref\ParkSuhN{J.-H. Park and Y. Suh, {\xit ``U-gravity: SL(N)''},
\jhep{13}{04}{2013}{102} [\arxiv{1402}{5027}].]}

\ref\HohmZwiebachLarge{O. Hohm and B. Zwiebach, {\xit ``Large gauge
transformations in double field theory''}, \arxiv{1207}{4198}.}

\ref\PalmkvistTensor{J. Palmkvist, {\xit ``The tensor hierarchy
algebra''}, \JMP{55}{2014}{011701} [\arxiv{1305}{0018}].}

\ref\CarboneCederwallPalmkvist{L. Carbone, M. Cederwall and
J. Palmkvist, {\xit ``Generators and relations for Lie superalgebras
of Cartan type''}, in preparation.}

\ref\BeyondEEleven{G. Bossard, A. Kleinschmidt, J. Palmkvist,
C.N. Pope and E. Sezgin, {\xit ``Beyond $E_{11}$''},
\jhep{17}{05}{2017}{020} [\arxiv{1703}{01305}].}

\ref\HitchinLectures{N. Hitchin, {``\xit Lectures on generalized
geometry''}, \arxiv{1010}{2526}.}

\ref\deWitNicolaiSamtleben{B.~de Wit, H.~Nicolai and H.~Samtleben,
{\xit ``Gauged supergravities, tensor hierarchies, and M-theory},
\jhep{08}{02}{2008}{044}
[\arxiv{0801}{1294}].}

\ref\KacSuperalgebras{V.~Kac, {\xit ``Lie superalgebras''}, Adv. Math. {\xbf26}
(1977) 8.}

\ref\KacFiniteGrowth{V.~Kac, {\xit ``Simple irreducible graded Lie algebras of finite growth''}, Math. USSR Izv. {\xbf2}
(1968) 1271.}

\ref\RayBook{U.~Ray, {\xit ``Automorphic forms and Lie
superalgebras''}, Springer
({\xold2006}).}

\ref\BossardKleinSchmidtLoops{G. Bossard and A. Kleinschmidt, {\xit
``Loops in exceptional field theory''}, \jhep{16}{01}{2016}{164} [\arxiv{1510}{07859}].}

\ref\HohmSamtlebenEhlers{O. Hohm and H. Samtleben, {\xit ``U-duality
covariant gravity''}, \jhep{13}{09}{2013}{080} [\arxiv{1307}{0509}].}

\ref\HohmMusaevSamtleben{O. Hohm, E.T. Musaev and H. Samtleben, {\xit
``$O(d+1,d+1)$ enhanced double field
theory''}, \jhep{17}{10}{2017}{086} [\arxiv{1707.06693}].}

\ref\BlumenhagenHasslerLust{R. Blumenhagen, F. Hassler and D. L\"ust,
{\xit ``Double field theory on group manifolds''},
\jhep{15}{02}{2015}{001} [\arxiv{1410}{6374}].}

\ref\BlumenhagenBosqueHasslerLust{R. Blumenhagen, P. du Bosque,
F. Hassler and D. L\"ust, 
{\xit ``Generalized metric formulation of double field theory on group
manifolds''}, \arxiv{1502}{02428}.}

\ref\HasslerTopology{F. Hassler, {\xit ``The topology of double field
theory''}, \arxiv{1611}{07978}.}

\ref\HasslerTopology{F. Hassler, {\xit ``Poisson--Lie T-duality in
double field theory''}, \arxiv{1707}{08624}.}


\def\tilde{\widetilde}


\line{
\epsfysize=18mm
\epsffile{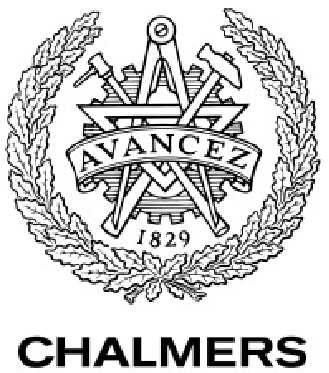}
\hfill}
\vskip-16mm

\line{\hfill}
\line{\hfill Gothenburg preprint}
\line{\hfill November, {\old2017}
}
\line{\hrulefill}


\headtext={Cederwall, Palmkvist: 
``Extended geometries''}

\vfill

\centerline{\sixteenhelvbold
Extended geometries}

%

%

\vfill

\centerline{\twelvehelvbold Martin Cederwall
and Jakob Palmkvist}

\vfill
\vskip-1cm

\centerline{\it Division for Theoretical Physics}
\centerline{\it Department of Physics}
\centerline{\it Chalmers University of Technology}
\centerline{\it SE 412 96 Gothenburg, Sweden}

\vfill

{\narrower\noindent \underbar{Abstract:}
We present a unified and completely general formulation of extended
geometry, characterised by a Kac--Moody algebra and a highest weight
coordinate module. Generalised diffeomorphisms are
constructed, as well as solutions to the section
constraint. Generically, additional (``ancillary'') gauge
transformations are present, and we give a concrete criterion
determining when they appear. A universal form of the
(pseudo-)action
determines the dynamics in all cases without ancillary
transformations, and also for a restricted set of cases based on the
adjoint representation of a finite-dimensional simple Lie group.
Our construction reproduces (the internal sector of)
all previously considered cases of double and
exceptional field theories.

\smallskip}
\vfill

\font\xxtt=cmtt6

\vtop{\baselineskip=.6\baselineskip\xxtt
\line{\hrulefill}
\catcode`\@=11
\line{email: martin.cederwall@chalmers.se, jakob.palmkvist@chalmers.se\hfill}
\catcode`\@=\active
}

\eject


\contentsout

\section\IntroSection{Introduction}During the last years, much research has been
devoted to 
models of gravity together with other fields,
obtained by extending $d$-dimensional space to some
module of a structure group $G$. The structure group plays an
important r\^ole for the gauge symmetries of these models, the generalised
diffeomorphisms. Locally, $d$-dimensional space and the group
$GL(d)\subset G\times \RR^+$ are recovered through a section
constraint, an algebraic condition on allowed momenta. In this paper, we let the term {\it extended
geometry} denote any such model.
The most important classes of extended geometry are double field
theory and exceptional field theory, motivated by dualities in string
theory.

Recently, universal expressions for the invariant tensors used in the
construction of extended geometries have been uncovered
[\PalmkvistBorcherds,\ENinePaper], pointing towards a unified
treatment. These expressions constitute a generalisation of a generic
identity fulfilled for elements in a minimal orbit, that has appeared
in the mathematics literature [\PetersonKac].
Although double field theory
[\Tseytlin\skipref\SiegelI\skipref\SiegelIII\skipref\HitchinLectures\skipref\HullT\skipref\HullDoubled\skipref\HullZwiebachDFT\skipref\HohmHullZwiebachI\skipref\HohmHullZwiebachII\skipref\HohmKwakFrame\skipref\HohmKwak\skipref\JeonLeeParkI\skipref\JeonLeeParkII\skipref\JeonLeeParkIII\skipref\HohmZwiebachGeometry\skipref\HohmKwakZwiebachII\skipref\JeonLeeParkSuh\skipref\JeonLeeParkRR\skipref\HohmZwiebachLarge\skipref\Park\skipref\BermanCederwallPerry\skipref\CederwallGeometryBehind\skipref\CederwallDuality\skipref\BlumenhagenHasslerLust--\BlumenhagenBosqueHasslerLust]
has a generic form, thanks to
the tensor formalism of $O(d,d)$, exceptional field
theory
[\HullM\skipref\PachecoWaldram\skipref\Hillmann\skipref\BermanPerryGen\skipref\BermanGodazgarPerry\skipref\BermanGodazgarGodazgarPerry\skipref\BermanGodazgarPerryWest\skipref\CoimbraStricklandWaldram\skipref\CoimbraStricklandWaldramII\skipref\BermanCederwallKleinschmidtThompson\skipref\ParkSuh\skipref\CederwallI\skipref\CederwallII\skipref\HohmSamtlebenEhlers\skipref\HohmSamtlebenI\skipref\HohmSamtlebenII\skipref\HohmSamtlebenIII\skipref\CederwallRosabal\CederwallExceptionalTwistors\skipref\BossardKleinSchmidtLoops--\HohmMusaevSamtleben]
has typically relied on a case by case treatment.

It has not been clear what the most general
situation is.
Some important examples of exceptional field theory for high rank
groups ($E_8$, $E_9$) exhibit the presence of additional constrained
local transformations
[\HohmSamtlebenEhlers,\HohmSamtlebenIII,\CederwallRosabal,\HohmMusaevSamtleben,\ENinePaper],
for which we use the term {\it ancillary transformations}.  However,
the classifications in refs. [\StricklandConstable,\Baraglia] rely on
the absence of such transformations, so it is relevant to reconsider
the general setting.

In the present paper, we develop a general formalism for extended
geometry based on a choice of a Kac--Moody algebra $\fg$ and an
irreducible highest weight module $R(\lambda)$. As our only
restrictions on the choice, we require the Cartan matrix of $\fg$ to
be indecomposable and symmetrisable, and the module $R(\lambda)$ to be
integrable (so that the Dynkin labels of the highest weight $\lambda$
are non-negative integers) [\KacBook].  All expressions for
generalised diffeomorphisms, section constraints etc., are universal,
and the formalism allows for a unified treatment of all extended
geometries. We limit ourselves to local properties; it is not clear if
the treatment of global issues may be facilitated by our methods.  Our
investigation is also limited to what in a compactification would be
the ``internal'' space, and the fields for which we formulate the
dynamics are those that from an ``external'' point of view are
scalars.

In Section \SectionDiffSection, the basic concepts of extended
coordinates, section constraints and generalised diffeomorphisms are
introduced.  We use the quadratic Casimir to obtain the general form
of the section constraint, and apply it in the construction of
generalised diffeomorphisms. The result shows that ancillary
transformations occur in many cases.  In Section \ExtAlgSection\ we
show how the general form of the section constraint may be derived
from bosonic and fermionic extensions of the structure algebra. This
alternative approach is then used in Section \AncillarySection, where
we give a concrete criterion in terms of the structure algebra and
coordinate module for whether ancillary transformations appear or not.
A pseudo-action, encoding the full dynamics for a limited set of
cases, is given in Section \DynamicsSection.  In
Section \ExamplesSection, examples are examined, some of which are
well known, and some new.  We conclude with a summary and outlook in
Section \ConclusionsSection.

\section\SectionDiffSection{Extended space, sections, and generalised
diffeomorphisms}We consider models based on a Kac--Moody algebra
$\fg$, exponentiated to a group $G$ (extending the structure group
$GL(d)$ in ordinary geometry), and an irreducible and integrable
highest weight module $R(\lambda)$ (extending the ordinary coordinate
module).  Derivatives will thus transform in the dual module
$\overline{R(\lambda)}$.

We denote the Cartan matrix of $\fg$ by $a_{ij}$ ($i,j=1,\ldots,r$) if
it is invertible. Otherwise, if the corank of the Cartan matrix is
$m\geq 1$ (in particular if $\fg$ is affine, $m=1$), then we extend
the range of indices to $i,j=1,\ldots,n$, where $n=r+m$, and let
$a_{ij}$ be an invertible $n\times n$ matrix obtained by adding $m$
rows and columns to the Cartan matrix (corresponding to a
``realisation'' of it [\KacBook]).  We will always assume that $a$ is
symmetrisable, which means that there is a diagonal matrix $d$ with
nonzero entries $d_j$ such that $(da)_{ij}$ is a symmetric
matrix. This gives rise to a non-degenerate symmetric bilinear form on
the Cartan subalgebra, and a corresponding metric on the weight
space.\foot{If the Cartan matrix is not invertible, $m\geq1$, then the
set of simple roots $\alpha_1,\ldots,\alpha_r$ in the weight space is
supplemented by $m$ additional basis elements
$\alpha_{r+1},\ldots,\alpha_n$, which strictly speaking are no
roots. However, for simplicity we will refer to all basis elements
$\alpha_1,\ldots,\alpha_n$ as simple roots.}  We use the conventions
that $d$ expresses half the lengths squared of the simple roots,
$d_i={(\alpha_i,\alpha_i)\over 2}$, and $da$ their mutual inner
products, $(da)_{ij}=(\alpha_i,\alpha_j)$.  Thus
$a_{ij}=(\alpha_i{}^\vee,\alpha_j)$, where
$\alpha_i{}^\vee={2 \alpha_i \over (\alpha_i,\alpha_i)}$ is the coroot
of $\alpha_i$.  It is often convenient to symmetrise $a$ from the
right with the inverse of $d$, so that the symmetrised matrix is $\hat
a_{ij}=(ad^{-1})_{ij}=(\alpha_i{}^\vee,\alpha_j{}^\vee)$.  With
weights expressed as $\lambda=\sum_{i}\lambda_i\Lambda_i$, where the
fundamental weights $\Lambda_i$ are given by
$(\Lambda_i,\alpha^\vee_j)=\delta_{ij}$, the weight space metric is
then given by $\hat a^{-1}$, so that $(\lambda,\lambda)=\sum_{ij}(\hat
a^{-1})^{ij}\lambda_i\lambda_j$. The Dynkin labels $\lambda_i$ of the
(highest) weights $\lambda$ that we consider are (non-negative)
integers.  We note that $d$ is not uniquely given by $a$, but only up
to an overall factor, and that all the entries have the same sign.
The (standard) convention is that the longest simple roots have
$(\alpha_i,\alpha_i)=2$.  Throughout the paper, we will take the real
form of $\fg$ to be the split one, \ie, the one associated with the
weight space decomposition.

Following ref. [\BermanCederwallKleinschmidtThompson], we consider
generalised diffeomorphisms of the form
$$
\LL_\xi V^M=\xi^N \partial_N V^M-\partial_N \xi^M V^N
+Y^{MN}{}_{PQ}\*_N\xi^PV^Q\komma\Eqn\GenDiffY
$$ 
where $Y$ is a $\fg$-invariant tensor that needs to satisfy certain
identities in order for the generalised diffeomorphisms to close into
an algebra and to be covariant with respect to themselves.  The first
of these identities is simply the section constraint
$$
Y^{MN}{}_{PQ}(\partial_M \otimes \partial_N)=0 \Eqn\GenSectionCondition
$$
and, as shown in ref. [\PalmkvistBorcherds], all but one of the other
identities can be conveniently combined into the single
``fundamental'' identity
$$
\eqalign{
(Z^{NT}{}_{SM}Z^{QS}{}_{RP}&-Z^{QT}{}_{SP}Z^{NS}{}_{RM}\cr
&-Z^{NS}{}_{PM}Z^{QT}{}_{RS}+Z^{ST}{}_{RP}Z^{NQ}{}_{SM})(\partial_N \otimes \partial_Q)
=0\;,} \Eqn\FundIdentity
$$
where $Z^{MN}{}_{PQ}=Y^{MN}{}_{PQ}-\delta^M_P\delta^N_Q$.

Already
demanding that the last two terms in the transformation (\GenDiffY)
represent a local transformation on $V$ in the structure algebra
$\fg\oplus\RR$ constrains $Y$ to the form
$$
Y^{MN}{}_{PQ} = -k \eta_{\alpha\beta} T^{\alpha N}{}_P T^{\beta M}{}_Q
+ \beta \delta^N_P\delta^M_Q+\delta^M_P\delta^N_Q\;, \Eqn\GenDiffAnsatz
$$
where $T^{\alpha N}{}_P$ are the representation matrices for
$R(\lambda)$ in a basis $T^\alpha$ of $\fg$ and $\eta_{\alpha\beta}$
is the inverse of the invariant bilinear form.  Then it is easily
checked that, for any values of the constants $k$ and $\beta$ in
(\GenDiffAnsatz), the condition (\FundIdentity) is automatically
satisfied if the section constraint (\GenSectionCondition) is. This
condition is however not sufficient for closure of the generalised
diffeomorphisms, but leads to a remaining term
$$
\big([{\scr L}_\xi,{\scr L}_\eta] -
{\scr L}_{{1\over2}({\scr L}_\xi \eta-{\scr L}_\eta \xi)}\big)V^M
=\fr2Z^{MP}{}_{QN}Y^{QR}{}_{ST}\xi^T \partial_P \partial_R \eta^SV^N
-(\xi\leftrightarrow\eta)\;,
\Eqn\CovarianceComputation  
$$
that vanishes in some cases (\eg, $\fg=E_r$, $r \leq 7$) but not in
others (\eg, $\fg={E}_8$).  As we will see later, whether it vanishes
or not does not depend on the constants $k$ and $\beta$, but is
entirely a property of the Lie algebra $\fg$ and the module
$R(\lambda)$.  What does determine the constants $k$ and $\beta$
(although not always fully) is the requirement that the ``extended
geometry'' indeed is an extension of ordinary geometry, in the sense
that we recover ordinary geometry when we solve the section constraint
(\GenSectionCondition). This means that the solutions (``sections'')
must be $d$-dimensional linear subspaces of $R(\lambda)$ related to
each other by rotations in $G$, and with a stability group containing
$GL(d)$, such that the generalised diffeomorphisms in a section reduce
to ordinary diffeomorphisms together with transformations of other
gauge fields.  We will look closer at this constraint below, after a
brief interlude on notation.

The relations above are written in tensor notation. However, we will
often use an index-free notation, where the derivatives and vector
fields are represented by bra and ket states, respectively.  The
advantage of using the bra--ket notation is, besides making many
equations less cluttered, that it goes well together with the standard
treatment of highest/lowest weight modules of Lie algebras.  Thus,
$\xi^M\leftrightarrow\ket\xi$ and $\*_M\leftrightarrow\bra\*$. The
section constraint (\GenSectionCondition) takes the index-free form
$$
\bra\*\otimes\bra\*Y=0\;,\Eqn\IndexFreeSectionCondition
$$
while the expression (\GenDiffAnsatz) for the $Y$ tensor reads
$$
\sigma Y=-k\eta_{\alpha\beta}T^\alpha\otimes T^\beta+\beta+\sigma\;,\eqn
$$
where $\sigma$ is the permutation operator, $\sigma\ket a\otimes\ket
b=\ket b\otimes\ket a$, or, on operators, $\sigma A\otimes B=B\otimes
A\,\sigma$.

The section constraint can be introduced in two steps.  The first one
(which sometimes goes under the name ``weak section constraint'')
demands that momenta (derivatives) lie in a minimal orbit under the
action of $G$. This is equivalent to the statement that the
symmetrised product $\*^2$ only contains the module
$\overline{R(2\lambda)}$, dual to the highest module in the tensor
product $\otimes^2R(\lambda)$. The term ``highest'' here refers to the
partial ordering of (highest) weights.  The solution of the weak
constraint is not a linear space, but a c\^one.

In the second step, we demand that the product of any two momenta only
contains the lowest symmetric and antisymmetric modules in the tensor
product.  While obviously the highest module in the symmetrised
product $\vee^2R(\lambda)$ is $R(2\lambda)$, the highest module in the
the antisymmetrised product $\wedge^2R(\lambda)$ is in general
reducible, and consists of the sum of all irreducible highest weight
modules $R(2\lambda-\alpha_i)$, where the simple roots $\alpha_i$ are
the ones with $\lambda_i=(\lambda,\alpha^\vee_i)
\neq0$. 
This is easily seen from an expansion of $R(\lambda)$, starting from
the highest weight state $\ket\lambda$, followed by the states
$f_i\ket\lambda=\ket{\lambda-\alpha_i}$ for $\lambda_i\neq0$.  The
highest antisymmetric highest weight states then are
$$
\ketket{2\lambda-\alpha_i}
=\ket\lambda\otimes\ket{\lambda-\alpha_i}-\ket{\lambda-\alpha_i}\otimes\ket\lambda
\;.\eqn
$$
As will soon be clear, not all of these should survive for a solution
to the section constraint, but only those with $\lambda_i=1$ and
${2\over(\alpha_i,\alpha_i)}=k$ for some given $k$. We denote the
corresponding index set by
$$
{\cal I}_k\subset\bigl\{i\,|\,\lambda_i=1,
\,{2\over(\alpha_i,\alpha_i)}=k\bigr\}\;.\eqn
$$ 
The section constraint will thus be equivalent to 
$$\Bigl.(\*\otimes\*)\Bigr\vert_{\overline{R_2\oplus\tilde
R_2}}=0\;,\Eqn\RTwoSectionConstr
$$
where
$$
\eqalign{
R_2&=\vee^2R(\lambda)\ominus R(2\lambda)\;,\cr
\tilde R_2&=\wedge^2R(\lambda)\ominus
\bigoplus\limits_{i\in{\cal I}_k}R(2\lambda-\alpha_i)\punkt
}\Eqn\RtwoDefinition
$$
The modules $R_2$ and $\tilde R_2$, which do not need to be
irreducible, have a natural interpretation in terms of extended
algebras, as explained in Section \ExtAlgSection.

As we will see, the constant $k$ in ${\cal I}_k$ is the same as the
one in eq. (\GenDiffAnsatz). To get an as sensible and less
restrictive model as possible, we set $k={2\over(\alpha_i,\alpha_i)}$
if all simple roots $\alpha_i$ have the same length squared
$(\alpha_i,\alpha_i)$, or if there is only one simple root $\alpha_i$
such that $\lambda_i=1$. If there is more than one simple root with
$\lambda_i=1$, and they have different lengths, then we set
$k={2\over(\alpha_i,\alpha_i)}$ for one of them (thus $k$ is not fully
determined by $\fg$ and $\lambda$ in this case).  If $\lambda_i\neq 1$
for all simple roots $\alpha_i$, then we let $k$ be such that $k\geq
{2\over(\alpha_i,\alpha_i)}$ for all simple roots $\alpha_i$ (\ie,
$k\geq 1$ with the convention that the longest simple roots $\alpha_i$
have $(\alpha_i,\alpha_i)=2$) and otherwise undetermined.

Let us investigate what the section constraint (\RTwoSectionConstr)
implies in terms of algebraic conditions.  We will use the quadratic
Casimir operator, which is well defined on highest weight modules of
any Kac--Moody algebra with a symmetrisable Cartan matrix.  It is
$$
C_2=\fr2\eta_{\alpha\beta}:T^\alpha T^\beta:+(h,\varrho)
=\sum\limits_{\alpha\in\Delta_+}e_{-\alpha}e_\alpha+\fr2(h,h)+(h,\varrho)
\;.\eqn
$$
Here, $\varrho$ is the Weyl vector, defined so that
$(\varrho,\alpha^\vee_i)=1$. The term $(h,\varrho)$ is a normal
ordering term, which for finite-dimensional algebras can be absorbed
into a symmetric ordering, since then
$\varrho=\fr2\sum_{\alpha\in\Delta_+}\alpha$.  Since $C_2$ takes the
same value on all elements in an irreducible module $R(\lambda)$, it
is enough to evaluate it on the highest weight state, where one gets
$$
C_2(R(\lambda))=\fr2(\lambda,\lambda+2\varrho)\;.\eqn
$$
From this expression we also immediately obtain
$$
\eqalign{
C_2(R(2\lambda))&=2C_2(R(\lambda))+(\lambda,\lambda)\;,\cr
C_2(R(2\lambda-\alpha_i))&=2C_2(R(\lambda))+(\lambda,\lambda)
-\lambda_i(\alpha_i,\alpha_i)\;,\cr }\eqn
$$
where $\alpha_i$ is any simple root such that
$\lambda_i=(\lambda,\alpha_i{}^\vee)>0$.

Now, consider a representative in the minimal orbit of $R(\lambda)$,
which can be chosen proportional to the highest weight state
$\ket\lambda$ itself.  Following Section {\old6}.{\old2} of
ref. [\ENinePaper], we then have
$$
\eqalign{
0&=\left[C_2(R(2\lambda))-2C_2(R(\lambda))-(\lambda,\lambda)\right] \ket\lambda\otimes\ket\lambda\cr
                  &=\left[\fr2\eta_{\alpha\beta}:T^\alpha
                  T^\beta:+(h,\varrho)\right]
\ket\lambda\otimes\ket\lambda
-\left(\left[\fr2\eta_{\alpha\beta}:T^\alpha
 T^\beta:+(h,\varrho)\right]
\ket\lambda\right)\otimes\ket\lambda\cr
&\qquad-\ket\lambda\otimes
\left[\fr2\eta_{\alpha\beta}:T^\alpha T^\beta:+(h,\varrho)\right]
\ket\lambda
-(\lambda,\lambda)\ket\lambda\otimes\ket\lambda\cr
&=\left[\eta_{\alpha\beta}T^\alpha\otimes
T^\beta-(\lambda,\lambda)\right]\ket\lambda\otimes\ket\lambda\;.\cr
}\Eqn\CasimirEqS
$$
Similarly, for $i$ such that $\lambda_i=1$, we obtain
$$
\eqalign{
0&=\left[C_2(R(2\lambda-\alpha_i))-2C_2(R(\lambda))-(\lambda,\lambda)
        +(\alpha_i,\alpha_i)\right] \ketket{2\lambda-\alpha_i}\cr
        &=\left[\eta_{\alpha\beta}T^\alpha\otimes
        T^\beta-(\lambda,\lambda)+(\alpha_i,\alpha_i)\right]\ketket{2\lambda-\alpha_i}
\punkt
}\Eqn\CasimirEqA
$$

Next, consider two vectors in a section. Without loss of generality,
take one of them to be $\ket\lambda$ and the other one $\ket
q=\sum_{\ell\geq0}\ket q_\ell$, with components given by
$(h,\lambda)\ket q_\ell=((\lambda,\lambda)-\ell)\ket q_\ell$ in the
grading induced by $\lambda$. If the right hand side of
eq. (\CasimirEqS) annihilates symmetric products of vectors in a
section, we must have
$$
\left[\eta_{\alpha\beta}T^\alpha\otimes
T^\beta-(\lambda,\lambda)\right]
\left(\ket\lambda\otimes\ket q+\ket
q\otimes\ket\lambda\right)=0\;.\Eqn\QLambdaEq
$$
Since the equation is linear in $\ket q$, we can treat the terms in
the grading decomposition separately. We get
$$
\eqalign{
&\left[\eta_{\alpha\beta}T^\alpha\otimes
T^\beta-(\lambda,\lambda)\right]
\left(\ket\lambda\otimes\ket q_\ell+\ket
q_\ell\otimes\ket\lambda\right)\cr
&\quad=(1+\sigma)\Biggl(\sum\limits_{\alpha\in\Delta_+} e_\alpha\ket
q_\ell\otimes e_{-\alpha}\ket\lambda -\ell\ket\lambda\otimes\ket
q_\ell
\Biggr)\;.\cr
}\eqn
$$
This shows that, for $\ell\geq1$, we must have $\ket q_\ell\propto
e_{-\alpha}\ket\lambda$, for some positive root $\alpha$. In order for
the sum to give a single term, we must have $e_{\alpha'}\ket q_\ell=0$
or $e_{-\alpha'}\ket\lambda=0$ for all positive roots
$\alpha'\neq\alpha$. The root $\alpha$ must thus be $\alpha_i+\beta$,
where $\alpha_i$ is a simple root with $\lambda_i\neq0$ and
$(\lambda,\beta)=0$.  Eq. (\QLambdaEq) is then satisfied if
$\ell=\lambda_i$.

We also need to verify that $\ket q$ itself satisfies the symmetric
constraint, \ie, that it lies in the minimal orbit.  This condition is
not linear, and terms with different degree may mix. Therefore, we
consider the maximal value of $\ell$.  For simplicity, we only display
the calculation for $\alpha=\alpha_i$. Then
$$
\eqalign{
&\left[\eta_{\alpha\beta}T^\alpha\otimes
T^\beta-(\lambda,\lambda)\right]
\ket q_\ell\otimes\ket q_\ell\cr
&\quad=\lambda_i\left[\ket\lambda\otimes f_i^2\ket\lambda
                 +f_i^2\ket\lambda\otimes\ket\lambda\right]
                 +\left[(\lambda-\alpha_i,\lambda-\alpha_i)-(\lambda,\lambda)\right]
                 f_i\ket\lambda\otimes f_i\ket\lambda\;,\cr }\eqn
$$
and thus $f_i\ket\lambda$ is in the minimal orbit only if
$f_i^2\ket\lambda=0$ and
$(\lambda-\alpha_i,\lambda-\alpha_i)=(\lambda,\lambda)$. Both
conditions give $\lambda_i=1$. The same holds for $\ket
q_1=e_{-\alpha_i-\beta}\ket\lambda$ as above.  This consideration
shows why the antisymmetric highest weight modules
$R(2\lambda-\alpha_i)$ are allowed to be non-vanishing if
$\lambda_i=1$. Note that this condition is obtained by demanding that
the symmetric constraint is satisfied by vectors in a linear subspace.

Continuation of this construction yields a linear subspace, a section,
of $R(\lambda)$ that behaves as a $GL(d)$ module. Both the symmetric
and antisymmetric products of vectors in the section contain a single
irreducible module, $R(2\lambda)$ and $R(2\lambda-\alpha_i)$,
respectively, depending on the choice of simple root with
$\lambda_i=1$.  The possible (representatives of) sections are
immediately read off from the Dynkin diagram: In addition to
$\ket\lambda$, one chooses a $f_i\ket\lambda$ with $\lambda_i=1$. Then
the section is sequentially enlarged with a neighbouring node (here
labelled $i+1$) as $f_{i+1}f_i\ket\lambda$ as long as
$\lambda_{i+1}=0$, and so on. If a branching in the Dynkin diagram is
encountered, one chooses one branch to proceed along. The process
stops at an end of the diagram, or before one encounters another node
$j$ with $\lambda_j\neq0$, or before one encounters a node with a
multiple connection.  This corresponds to choosing a maximal (\ie,
non-extendable) line of $d$ simply connected nodes of equal length in
the extended Dynkin diagram (see Section \ExtAlgSection), an
``extended gravity line''.

From this, it follows that a representative of the section is
identified as the states in $R(\lambda)$ at degree $0$ in the grading
with respect to the node (or one of the nodes) lying next to the
gravity line in the Dynkin diagram. Call this node number $j$. The
degree of a weight $\mu$ in this grading is given by ${2 \over
(\alpha_j,\alpha_j)}(\Lambda_j,\mu)$. The representative section is
specified by $\Lambda_j$, and other sections by elements in the orbit
of $\Lambda_j$. It follows that a section is specified by an element
in the minimal orbit of $\Lambda_j$
[\BermanCederwallKleinschmidtThompson], \ie, a $\phi\in R(\Lambda_j)$
such that $\phi^2\in R(2\Lambda_j)$. A familiar example is the
parametrisation of isotropic subspaces in the fundamental
representation of $D_r$ (sections in double field theory) in terms of
pure spinors.

All vectors $\ket p$, $\ket q$ in a section thus satisfy the universal
constraint
$$
Y\ket p\otimes\ket q=0\;,\eqn
$$
where
$$
\sigma Y=k\left[-\eta_{\alpha\beta}T^\alpha\otimes T^\beta
+(\lambda,\lambda)\right]+\sigma-1\;,\Eqn\SigmaY
$$
which is (the index-free version of) the expression (\GenDiffAnsatz)
with $\beta=k(\lambda,\lambda)-1$.  In all situations where there is a
section of dimension larger than 1, the constant $k$ has the fixed
value $k={2\over(\alpha_i,\alpha_i)}$, where $\lambda_i=1$.

Note that it is enough to consider the symmetric part of the
condition, the antisymmetric is then automatically satisfied (for a
given choice of section).  The $Y$ tensor commutes with $\sigma$,
which means that it can be decomposed as
$Y^{MN}{}_{PQ}=Y^{(MN)}{}_{(PQ)}+Y^{[MN]}{}_{[PQ]}$. The section
constraint is imposed on derivatives as in
eq. (\IndexFreeSectionCondition).

In ref. [\ENinePaper] this necessary form of the $Y$ tensor was shown
for the cases where $\lambda$ is a fundamental weight dual to a coroot
to a long simple root (\ie, $(\alpha_i,\alpha_i)=2$); the present
treatment holds for arbitrary Kac--Moody groups with a symmetrisable
Cartan matrix and arbitrary coordinate modules $R(\lambda)$. Many
cases seem quite uninteresting. Any highest weight with all
$\lambda_i\neq1$ leads to a 1-dimensional section, spanned by
$\ket\lambda$. Most of these should be unphysical or uninformative. An
exception is $SL(2)$ with the adjoint as coordinate module. This
corresponds to the Ehlers symmetry arising on reduction of gravity
from 4 to 3 dimensions, and the section should indeed be
1-dimensional.  It also seems like ``gravity lines'' containing short
roots ($(\alpha_i,\alpha_i)<2$) typically do not correspond to
interesting situations. The typical example would be when the
coordinate module is the fundamental of an algebra $C_r$. This example
is mentioned in Section \ExamplesSection.  In
Section \AncillarySection, we will further investigate the
consequences of different possible choices.

Let us now come back to the generalised diffeomorphisms (\GenDiffY),
which in index-free notation take the form
$$
\LL_\xi\ket V=\bra{\*_V}\xi\rangle\otimes\ket V
+\bra{\*_\xi}(-k\eta_{\alpha\beta}T^\alpha\otimes
T^\beta+k(\lambda,\lambda)-1)\ket\xi\otimes\ket V\punkt
\eqn
$$

Commuting two such transformations leads to the remainder term
(\CovarianceComputation) (antisymmetrised in $\xi$ and $\eta$), which
after some standard calculation becomes
$$
\left([\LL_\xi,\LL_\eta]-\LL_{{1\over2}(\LL_\xi\eta-\LL_\eta\xi)}\right)\ket V
=\Sigma_\alpha T^\alpha\ket V\komma\Eqn\RemainderTerm
$$
where the element $\Sigma=\Sigma_\alpha T^\alpha\in\fg$ is given as
$$
\eqalign{
\Sigma^\alpha&=\Fr k2\bra{\*_\eta}\otimes\bra{\*_\eta}S^\alpha
\ket\xi\otimes\ket\eta-(\xi\leftrightarrow\eta)\;,\cr
\hbox{with}\;\;S^\alpha&=-kf^\alpha{}_{\beta \gamma}T^\beta\otimes T^\gamma
+T^\alpha\otimes1-1\otimes T^\alpha}
\Eqn\SigmaEq
$$
(adjoint indices raised and lowered with $\eta$).  The operator
$S^\alpha$ has the symmetry $\sigma S^\alpha=-S^\alpha\sigma$, meaning
that it decomposes as $S^{\alpha MN}{}_{PQ}=S^{\alpha
(MN)}{}_{[PQ]}+S^{\alpha [MN]}{}_{(PQ)}$. Only the first term
contributes in $\Sigma^\alpha$.  Without indices it can be written
${1+\sigma\over2}S^\alpha=S^\alpha{1-\sigma\over2}$, and obtained as
$$
\Fr{1+\sigma}2S^\alpha=\Fr{1+\sigma}2[1\otimes
T^\alpha,Y]\punkt\Eqn\SigmaInTermsOfY
$$
Therefore, it follows that (in $\Sigma^\alpha$) the operator
$S^\alpha$ may be replaced by
$$
S^\alpha\rightarrow(1\otimes T^\alpha)Y_-\;,\Eqn\SAlphaEq
$$
where $Y_-=Y{1-\sigma\over2}$.

We see that, in the generic situation, the generalised diffeomorphisms
will not close among themselves when acting on a vector.  There will
be additional transformations present, which are local transformations
in $\fg$ of a restricted type. We use the term ``ancillary
transformations'' for these extra gauge symmetries.  Such
transformations have already been shown to be important in a number of
situations [\HohmSamtlebenEhlers,\HohmSamtlebenIII,\CederwallRosabal,\HohmMusaevSamtleben,\ENinePaper].
Eq. (\SigmaEq) provides a very simple expression for the generated
ancillary transformation.  The usual (previous) form of the extra
remaining term on the right hand side of eq. (\RemainderTerm) contains
the $Y$ tensor quadratically as $YZ$ (see
eq. (\CovarianceComputation)).  The new form (\SigmaEq) is only
quadratic in generators, and should be much easier to deal with. It
will appear naturally in the variations of the action of
Section \DynamicsSection.

We will examine the remainder term closer in
Section \AncillarySection, where a simple criterion for its presence
or absence will be given. The derivation relies in part on the concept
of extended algebras, which are introduced in the following Section.

\section\ExtAlgSection{Extended algebras}From the coordinate module $R(\lambda)$
as $R_1$ and ${\tilde R}_1$, and from $R_2$ and ${\tilde R}_2$ defined
in (\RtwoDefinition), representations $R_p$ and ${\tilde R}_p$ can be
defined for all positive integers $p$ (possibly trivial for all but
finitely many $p$) by extending the Lie algebra $\fg$ in a certain
way, and then decomposing the adjoint representation of the extended
algebra under $\fg$.  In the case of exceptional geometry, the
sequence $R_p$ was in ref. [\BermanCederwallKleinschmidtThompson]
shown to encode the infinite reducibility of the generalised
diffeomorphisms, and it agrees with tensor hierarchies in gauged
supergravity and exceptional field theory
[\deWitNicolaiSamtleben,\HohmSamtlebenI\skipref\HohmSamtlebenII--\HohmSamtlebenIII].

The form of the $Y$ tensor in exceptional geometry was constructed
from the extended algebras in ref. [\PalmkvistBorcherds].  In this
Section we review the construction, but for the general Kac--Moody
algebra $\fg$ rather than explicitly for ${E}_r$. The generalisation
from ${E}_r$ to $\fg$ is straightforward as long as the Cartan
matrices of $\fg$ and of the fermionic and bosonic extensions ${\scr
A}$ and ${\scr B}$ below are invertible. We will assume this until the
end of the Section, where we briefly describe the general case.

First we extend the Cartan matrix $a_{ij}$ ($i,j=1,\ldots,r$) of $\fg$
to the Cartan matrix $B_{IJ}$ ($I,J=0,1,\ldots,r$) of a contragredient
Lie superalgebra [\KacSuperalgebras] ${\scr B}$ by another row and
column such that
$$
B_{00}=0\;, \qquad B_{i0} = -\lambda_i\;, \qquad B_{ij}=a_{ij}\;,\eqn
$$
and such that $(DB)_{IJ}$ is a symmetric matrix, where $D$ is a
diagonal matrix with entries $D_0=1/k$ and $D_i=d_i$.  Thus
$B_{0i}={kd_i} B_{i0} = -{kd_i} \lambda_i$.

In the construction of ${\scr B}$ from the Cartan matrix $B$ one
starts with the Lie superalgebra generated by two odd elements
$e_0,f_0$ and $3r+1$ even elements $e_i,f_i,h_I$ modulo the relations
$$
[h_I,e_J]=B_{IJ}e_J\;, \qquad\quad [h_I,f_J]=-B_{IJ}f_J\;,\qquad
[e_I,f_J]=\delta_{IJ}h_J\;,\Eqn\ChevRel
$$
and factors out the maximal ideal that intersects the Cartan
subalgebra (spanned by the $h_I$) trivially.  In this case the
resulting contragredient Lie superalgebra ${\scr B}$ is a Borcherds
superalgebra, and the ideal is generated by the additional (Serre)
relations [\RayBook]
$$
[e_0,e_0]=[f_0,f_0]=0, \qquad i\neq J \ \Rightarrow\ ({\rm ad\,}
e_i)^{1-B_{iJ}}(e_J) = ({\rm ad\,}
f_i)^{1-B_{iJ}}(f_J)=0\;.\Eqn\SerreB
$$

The Lie superalgebra $\scr B$ can be decomposed as ${\scr B}
= \bigoplus_{p \in \mathbb{Z}} {\scr B}_p$, where $e_0 \in {\scr B}_1$
and $f_0 \in {\scr B}_{-1}$, and all other generators belong to ${\scr
B}_0$.  This is a (consistent) $\mathbb{Z}$-grading, which means
$[{\scr B}_p,{\scr B}_q]\subseteq {\scr B}_{p+q}$.  The even
subalgebra ${\scr B}_0$ is (as a Lie algebra) the direct sum of $\fg$
and a one-dimensional center spanned by an element $c$.  With a
normalisation such that $[c,e_0]=e_0$, the components of $c$ in the
basis $h_I$ are given by $c=\sum_I (B^{-1})^{0I}h_I$.

The subspaces ${\scr B}_{\pm 1}$ are irreducible $\fg$-modules under
the adjoint action of ${\scr B}_0$, and $f_0$ is a highest weight
vector of ${\scr B}_{-1}$ since $[e_i,f_0]=0$. The Dynkin labels are
given by
$$
[h_i,f_0]=-B_{i0}f_0 =\lambda_i f_0\;. \eqn
$$
Thus $\fg$ acts on ${\scr B}_{-1}$ and ${\scr B}_{1}$ in the
representations $R(\lambda)$ and $\overline{R(\lambda)}$,
respectively.  With bases $E^M$ and $F_M$ of ${\scr B}_1$ and ${\scr
B}_{-1}$, respectively, this means
$$
[T^\alpha,E^M]=-T^{\alpha M}{}_N E^N\;, \qquad
[T^\alpha,F_M]=T^{\alpha N}{}_M F_N\;.
\eqn
$$
In general, we denote the representation of $\fg$ corresponding to
${\scr B}_{-p}$ by $R_p$. Thus $R_1=R(\lambda)$, and it follows from
the Serre relations (\SerreB), which set the highest weight vector
$[f_0,f_0]$ of $R(2\lambda)$ to zero, that $R_2=\vee^2
R(\lambda) \ominus R(2\lambda)$.

The additional row and column in the Cartan matrix correspond to an
additional (odd) simple root $\beta_0$ in an extended weight space
with metric given by $(DB)_{IJ}=(\beta_I,\beta_J)$, where
$\beta_i=\alpha_i$. In particular $\beta_0$ is a null rot,
$(\beta_0,\beta_0)=0$. The corresponding invariant bilinear form on
$\scr B$ is given by $(h_I,h_J)=(\beta_I{}^\vee,\beta_J{}^\vee)$ on the Cartan subalgebra, where $\beta_0{}^\vee=k\beta_0$.
It is not symmetric on the whole of $\scr B$, but has a $\mathbb{Z}_2$-graded symmetry, consistent with the $\mathbb{Z}_2$-grading of $\scr B$.
In particular, $(e_0,f_0)=-(f_0,e_0)=k$. We choose a relative normalisation of the bases of ${\scr B}_1$ and ${\scr B}_{-1}$ such that
$(E^M,F_N)=-(F_N,E^M)=\delta^M_N$. On $\fg$ we have $(T^\alpha,T^\beta)=\eta^{\alpha\beta}$ as before. The length squared of the element $c$ is 
$(c,c)=k(B^{-1})^{00}$. 
Now we have
$$
[E^M,F_N]=
-\eta_{\alpha\beta} 
T^{\alpha M}{}_N T^\beta
+ {1\over k(B^{-1})^{00}} \delta^M_N c \eqn
$$
and
$$
f^M{}_N{}^P{}_Q \equiv ([[E^M,F_N],E^P],F_Q)= 
\eta_{\alpha\beta}T^{\alpha M}{}_NT^{\beta P}{}_Q + {1\over k(B^{-1})^{00}}\delta^M_N\delta^P_Q\;.\eqn
$$

Consider now the matrix $A_{IJ}$ given by $A_{00}=2$, and $A_{IJ}=B_{IJ}$ otherwise, \ie, if not $I=J=0$.
In the same way as the contragredient Lie superalgebra
$\scr B$ is constructed from $B$, we can construct a contragredient Lie algebra [\KacFiniteGrowth]
$\scr A$ from $A$. We thus replace $B_{IJ}$ by $A_{IJ}$ in the relations (\ChevRel), and let all generators be even.
For example, if $\fg={E}_r$, and $\lambda$ is the highest weight of the coordinate module in exceptional geometry,
then ${\scr A}$ is the Kac--Moody algebra ${E}_{r+1}$.

Similarly to ${\scr B}$, the Lie algebra $\scr A$ can be decomposed as ${\scr A} = \bigoplus_{p \in \mathbb{Z}} {\scr A}_p$,
where, as $\fg$-modules, ${\scr A}_p$ is isomorphic to ${\scr B}_p$ for $p=0,\pm1$. However, since we need to distinguish them from each other
we denote the basis elements of ${\scr A}_1$ and ${\scr A}_{-1}$ by $\tilde{E}^M$ and $\tilde{F}_M$, respectively. We also denote the
generators of $\scr A$ corresponding to $e_0$ and $f_0$
in $\scr B$ by ${\tilde e}_0$ and ${\tilde f}_0$. For the other generators $e_i,f_i,h_I$ there is no need to make this distinction. We thus have an isomorphism
${\scr B}_{\pm 1} \to {\scr A}_{\pm 1}$ mapping a general element
$U=U_M E^M$ to ${\tilde U}=U_M {\tilde E}^M$, and $V=V^M F_M$ to ${\tilde V}=V^M {\tilde F}_M$.
Note that whereas the elements $U,V$ are odd (fermionic) in the Lie superalgebra ${\scr B}$, the corresponding elements
${\tilde U},{\tilde V}$ are ordinary even (bosonic) elements in the Lie algebra $\scr A$. Accordingly,
the invariant bilinear form is now symmetric, like in $\fg$, so that $({\tilde E}^M,{\tilde F}_N)=({\tilde F}_N,{\tilde E}^M)=\delta^M_N$.

In the same way as for $\scr B$ we can now compute
$$
[{\tilde E}^M,{\tilde F}_N]= 
-\eta_{\alpha\beta} 
T^{\alpha M}{}_N T^\beta
+ {1\over k(A^{-1})^{00}} \delta^M_N c \eqn
$$
and
$$
{\tilde f}^M{}_N{}^P{}_Q \equiv ([[{\tilde E}^M,{\tilde F}_N],{\tilde E}^P],{\tilde F}_Q)= 
\eta_{\alpha\beta}T^{\alpha M}{}_NT^{\beta P}{}_Q + {1\over k(A^{-1})^{00}}\delta^M_N\delta^P_Q\;.\Eqn\Tildef
$$
We also define tensors
$$
g^{MN}{}_{PQ}=([E^M,E^N],[F_P,F_Q])\;, \qquad {\tilde g}^{MN}{}_{PQ}=([{\tilde E}^M,{\tilde E}^N],[{\tilde F}_P,{\tilde F}_Q])\;,\eqn
$$
which, by invariance of the bilinear form and the Jacobi identity, can be related to (anti-) symmetrisations of $f$ and $\tilde f$ as
$$
g^{MN}{}_{PQ}=-2f^{(M}{}_P{}^{N)}{}_Q\;, \qquad 
\tilde{g}^{MN}{}_{PQ}=2\tilde{f}^{[M}{}_P{}^{N]}{}_Q\;.\Eqn\fgrelation
$$
The weight $\lambda$ is an element in the weight space of the original Kac--Moody algebra $\fg$, and 
can be written in the basis of simple roots as 
$$
\lambda = \sum_j {(B^{-1})^{j0} \over
(B^{-1})^{00}} \alpha_j \quad \Rightarrow \quad
(\alpha_i{}^\vee,\lambda)=\sum_j B_{ij}{(B^{-1})^{j0} \over
(B^{-1})^{00}}=-B_{i0}=\lambda_i\;.
\eqn
$$
For its length squared we then get
$$
\eqalign{(\lambda,\lambda) &= \sum_{ij} d_i B_{ij} {(B^{-1})^{i0} \over (B^{-1})^{00}} {(B^{-1})^{j0} \over (B^{-1})^{00}} 
= -\sum_i d_{i} B_{i0} {(B^{-1})^{i0} \over (B^{-1})^{00}} \cr
&= - \sum_i {B_{0i} (B^{-1})^{i0} \over k(B^{-1})^{00}} = - {1\over k(B^{-1})^{00}}\;. 
} \Eqn\LambdalangdB
$$
Also $(A^{-1})^{00}$ can be related to $(B^{-1})^{00}$, by
$$
{1 \over (A^{-1})^{00}} = {\det A \over \det a} = {2\det a + \det B \over \det a}= 2 + {\det B \over \det a} = 2 + {1 \over (B^{-1})^{00}}\;. \Eqn\ABrelation
$$
Inserting (\ABrelation) into (\Tildef), and using (\LambdalangdB) and (\fgrelation) now gives
$$
{k\over2}(g^{MN}{}_{PQ}-{\tilde g}^{MN}{}_{PQ})= -k\eta_{\alpha\beta}T^{\alpha M}{}_PT^{\beta N}{}_Q
+k(\lambda,\lambda)\delta^M_P\delta^N_Q - \delta^M_P\delta^N_Q
+ \delta^N_P\delta^M_Q\;, \eqn
$$
or, in index-free notation,
$$
{k\over2}(g-\tilde g)=-k\eta_{\alpha\beta}T^\alpha\otimes T^\beta
+k(\lambda,\lambda)-1+\sigma\;.\eqn
$$
We see that this expression has the form (\GenDiffAnsatz), and agrees with eq. (\SigmaY). 
Thus the $Y$ tensor can be derived as 
$$
\sigma Y={k\over2}(g-\tilde g)\;.\Eqn\YFromg
$$

In the general case, when possibly any of the involved Cartan matrices is not invertible, it is convenient to go one step further and extend the (invertible)
matrix $a_{ij}$ ($i,j=1,\ldots,n$, where $n\geq r$, see the beginning of Section \SectionDiffSection)
by {\it two} rows and columns to a symmetrisable matrix $C_{IJ}$ ($I,J=-1,0,1,\ldots,n$) such that
$$
C_{(-1)0}=C_{0(-1)}=1\;,
\qquad C_{i0}=-\lambda_i\;, 
\qquad C_{0i}=-kd_i\lambda_i\;, 
\qquad C_{ij}=a_{ij}\;,
\eqn
$$
and all other entries are zero.
Note that $\det C = - \det a$. Let ${\scr C}$ be the Lie superalgebra constructed from $C$ in the same way as $\scr B$ is constructed from
$B$, with odd generators $e_{-1},e_0,f_{-1},f_0$ and even generators $e_i,f_i,h_j$, where $i=1,\ldots,r$ and $j=1,\ldots,n$.
Like $\scr B$, it can be decomposed as ${\scr C}=\bigoplus_{p \in \mathbb{Z}}{\scr C}_p$,
where $e_0 \in {\scr C}_1$ and $f_0 \in {\scr C}_{-1}$ and all other generators belong to ${\scr C}_0$. However, this is not a consistent $\mathbb{Z}$-grading;
${\scr C}_{\pm1}$ is the direct sum of
an even and an odd subspace, which can be identified with ${\scr A}_{\pm1}$ and ${\scr B}_{\pm1}$, respectively, by 
${\tilde E}^M=-[e_{-1},E^M]$ and ${\tilde F}_M=[f_{-1},F_M]$.
We then get
$$
\eqalign{
([[E^M,F_N],E^P],F_Q)&=\eta_{\alpha\beta}T^{\alpha M}{}_NT^{\beta P}{}_Q - {(C^{-1})_{(-1)(-1)}\over k}\delta^M_N\delta^P_Q\;,\cr
([[{\tilde E}^M,{\tilde F}_N],{\tilde E}^P],{\tilde
F}_Q)&=\eta_{\alpha\beta}T^{\alpha M}{}_NT^{\beta P}{}_Q +
{2-(C^{-1})_{(-1)(-1)}\over k}\delta^M_N\delta^P_Q\;,\cr}
\eqn
$$
and
similarly to eq. (\LambdalangdB) we have
$$
\eqalign{
{(C^{-1})^{(-1)(-1)} \over k} &= - \sum_i {C_{0i} (C^{-1})^{i(-1)} \over k(C^{-1})^{0(-1)}} = - \sum_i d_{i} C_{i0} {(C^{-1})^{i(-1)} \over (C^{-1})^{0(-1)}}\cr
&=  \sum_{ij} d_i C_{ij} {(C^{-1})^{i(-1)} \over (C^{-1})^{0(-1)}} {(C^{-1})^{j(-1)} \over (C^{-1})^{0(-1)}} = (\lambda,\lambda) \;.
} \Eqn\LambdalangdComvant
$$
Thus (\YFromg) holds also in the general case.

In Figure {\ExtAlgFigure } the Dynkin diagrams of the extended algebras $\scr A$, $\scr B$ and $\scr C$ are displayed, where the odd simple roots of zero length squared are represented by ``gray'' nodes. It is obvious from the figure that $\scr B$ is a subalgebra of $\scr C$. By performing an ``odd reflection'' with respect to the outermost gray node in the Dynkin diagram of $\scr C$ one can obtain an equivalent Dynkin diagram, where instead the embedding of $\scr A$ into $\scr C$ is
manifest [\PalmkvistBorcherds].

\vskip6\parskip
\vtop{\centerline{\epsffile{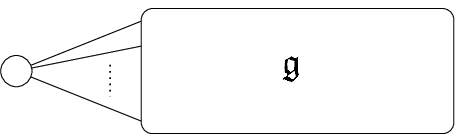}}
\vskip6\parskip
\centerline{\epsffile{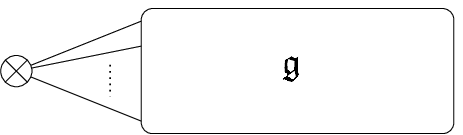}}
\TextFigure\ExtAlgFigure{\hskip-36pt\epsffile{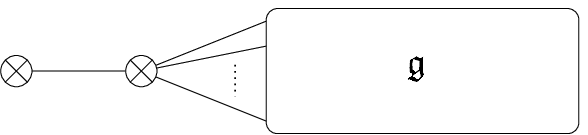}}{Dynkin
diagrams for the extended (super)algebras $\cal A$, $\cal B$ and
$\cal C$. The box represents the
Dynkin diagram of $\fg$.}}

\section\AncillarySection{Ancillary transformations}We
will now show how the formalism of ref. [\PalmkvistBorcherds],
reviewed and generalised in Section \ExtAlgSection, can be used to
determine in which 
cases the ancillary transformations appear or not.
As explained in the end of Section \SectionDiffSection,
the ancillary transformations vanish if and only if
the operator ${1 + \sigma \over 2} (1 \otimes T^\alpha)Y_-$ does when it acts from the right on derivatives.
With indices, using the results in the preceding Section, it can be written
$$
\eqalign{
\big( \small{1 + \sigma \over 2} (1 \otimes T^\alpha)Y_- \big){}^{MN}{}_{PQ}&=T^{\alpha(M}{}_R \delta^{N)}{}_S Y_-^{RS}{}_{PQ}
={k \over 2}T^{\alpha(M}{}_R \delta^{N)}{}_S {\tilde g}^{RS}{}_{PQ}\cr
&={k \over 2}([[T^\alpha,{\tilde E}^{(M}],{\tilde E}^{N)}],[{\tilde F}_P,{\tilde F}_Q])\;.
}\eqn
$$
If this expression 
is nonzero, 
and the indices $M$ and $N$ have any part in $R(2\lambda)$ for any $T^\alpha$,
then it will not vanish when contracted with two derivatives,
since the section constraint only removes the part in $R_2 = \vee^2 R(\lambda)\ominus R(2\lambda)$.
Since indices $M$ and $N$ of the anticommutator $[E^M,E^N]$ in ${\scr B}$
are automatically projected on $R_2$, a necessary and sufficient condition  
for the presence of 
ancillary transformations is the existence of an element $x \in \fg$ and
a symmetric tensor $X_{MN}$ such that $X_{MN}[E^M,E^N]=0$
but $X_{MN}[{\tilde E}^M,[{\tilde E}^N,x]]=0$, or equivalently, a set $\scr S$ of pairs $(U,V)$ of elements $U,V \in {\scr B}_{1}$
such that
$$
\sum_{(U,V)\in {\scr S}}[U,V]=0 \qquad {\rm and} \qquad
\sum_{(U,V)\in {\scr S}} \bigg([{\tilde U},[{\tilde V},x]]+[{\tilde V},[{\tilde U},x]]\bigg)\neq0 \Eqn\AncCondition
$$
in ${\scr B}_2$ and ${\scr A}_2$, respectively.

Since $[e_0,e_0]$ is always zero in $\scr B$, it is sufficient
to find 
an $x$
such that $[\tilde{e}_0,[\tilde{e}_0,x]] \neq 0$.
If there is a positive root $\alpha$ of $\fg$ such that
$(\alpha_0{}^\vee,\alpha) < -1$, then this condition is satisfied by
a corresponding root vector, $x=e_\alpha$, since
$$
[{\tilde f}_0,[{\tilde e}_0,e_{\alpha}]]=-[{h}_0,e_{\alpha}]=-(\alpha_0{}^\vee,\alpha) e_{\alpha}\neq0\;,\eqn
$$
implying that $[{\tilde e}_0,e_{\alpha}] \neq 0$, and then
$$
[{\tilde f}_0,[{\tilde e}_0,[{\tilde e}_0,e_{\alpha}]]]=
-[{\tilde h}_0,[{\tilde e}_0,e_{\alpha}]]
-[{\tilde e}_0,[{\tilde h}_0,e_{\alpha}]] =
(-2-2(\alpha_0{}^\vee,\alpha) )[{\tilde e}_0,e_{\alpha}]\neq 0\;,
\eqn
$$
implying that $[{\tilde e}_0,[{\tilde e}_0,e_{\alpha}]]\neq 0$.
To find a root $\alpha$ such that $(\alpha_0{}^\vee,\alpha)<-1$, we set $\alpha= \sum_i a_i \alpha_i$ so that
$$
(\alpha_0{}^\vee,\alpha)=\sum_i (\alpha_0{}^\vee,\alpha_i)a_i
=\sum_i k{(\alpha_i,\alpha_i)\over2}(\alpha_i{}^\vee,\alpha_0)a_i
=-\sum_i k{(\alpha_i,\alpha_i)\over2}\lambda_ia_i\,.
\eqn
$$
If $\lambda$ is not a fundamental weight (that is, if $\sum_i\lambda_i\geq2$), then 
$-\sum_i k{(\alpha_i,\alpha_i)\over2}\lambda_ia_i < -a_j$ for some $j$ such that $\lambda_j\geq1$,
and we can choose $\alpha=\alpha_j$.
Thus ancillary
transformations appear in this case.

Let us now assume that $\lambda$
is a fundamental weight, $\lambda_i=\delta_{ij}$ for some $j$. Then we have $-\sum_i k{(\alpha_i,\alpha_i)\over2}\lambda_ia_i = -a_j$, and
ancillary
transformations will again appear if there is a root $\alpha= \sum_i a_i \alpha_i$ of $\fg$
with $a_j\geq 2$. This includes
all infinite-dimensional cases. For finite-dimensional $\fg$, we can consider
the highest root $\theta= \sum_i c_i \alpha_i$
where $c_i$ are the Coxeter labels, and it follows that
if the
ancillary transformations are absent, then we must have $c_j=1$.
Conversely, if $c_j=1$, then $[{\tilde e}_0,[{\tilde e}_0,e_\theta]]=0$, and by the adjoint action of the raising operators $e_i$
(which commute with $e_\theta$) we get
$X_{MN}[{\tilde E}^M,[{\tilde E}^N,e_\theta]]=0$ for all symmetric tensors $X_{MN}$ such that
$X_{MN}[E^M,E^N]=0$, since these correspond to an irreducible representation with lowest weight $-2\lambda$.
Acting with the lowering operators $f_i$ we can then step down again from $e_\theta$ to any $x \in \fg$ and show that
$X_{MN}[{\tilde E}^M,[{\tilde E}^N,x]]=0$. Thus the condition $c_j=1$
is not only necessary for the absence of
ancillary transformations, but also sufficient.
(This can also be shown by studying the involved tensor product decompositions in all cases with $\lambda=\Lambda_j$ and $c_j=1$.
In many cases, ${\scr A}_2$ is zero- or one-dimensional, and it follows immediately that there is no room for ancillary transformations.)

We conclude that

{\narrower\noindent{\it the ancillary transformations vanish if and only
if $\fg$ is finite-dimensional, $\lambda$ is a fundamental weight
$\Lambda_j$, and the corresponding Coxeter label $c_j$ is equal to
$1$.}\smallskip}

Note that this can never happen at a node corresponding to a short
root, since the Coxeter label of a short root $\alpha_i$ always is
larger than 1.
The complete list of situations where ancillary transformations are
absent is thus:
\item{$\bullet$}{$\fg=A_r$, $\lambda=\Lambda_p$, $p=1,\ldots,r$
($p$-form representations);}
\item{$\bullet$}{$\fg=B_r$, $\lambda=\Lambda_1$ (the vector representation);}
\item{$\bullet$}{$\fg=C_r$, $\lambda=\Lambda_r$ (the 
symplectic-traceless $r$-form representation);}
\item{$\bullet$}{$\fg=D_r$,
$\lambda=\Lambda_1,\Lambda_{r-1},\Lambda_r$ (the vector and spinor
representations);}
\item{$\bullet$}{$\fg=E_6$, $\lambda=\Lambda_1,\Lambda_5$ (the
fundamental representations);}
\item{$\bullet$}{$\fg=E_7$, $\lambda=\Lambda_1$ (the fundamental
representation).}

The commutator between a generalised diffeomorphism with parameter
$\xi$ and an ancillary transformation with parameter $\Sigma=\Sigma_\alpha T^\alpha$
is trivially given by an ancillary transformation with parameter
$\LL_\xi\Sigma$.  Commuting two ancillary transformations does
not manifestly give a new transformation of the form (\SigmaEq). It is
however easy to give another argument for their closure. In at least one of the possible
section-adapted gradings (see Section \SectionDiffSection), the highest root vectors in ${\scr A}_{-2}$
appear at degree $-p$, where $p>0$. Thus the lower indices in
$Y_-^{RS}{}_{PQ}={k \over 2}{\tilde g}^{RS}{}_{PQ}$ correspond to degree $-p$ or lower. Since at the same time, by the section constraint,
derivatives are nonzero only at degree 0,
the degree of the parameter $\Sigma$ is also $-p$ or lower,
and the commutator will be of degree $-2p$ or lower.
We will see examples of this in Section 
\ExamplesSection.

\section\DynamicsSection{Dynamics}We now want to investigate if it is
possible to write a pseudo-action for fields in $G/K\times\RR^+$.
Due to the section constraint,
we do not yet consider the ``actions'' obtained here and earlier as
proper actions, unless integration is performed over some specified
section. They provide, however, an efficient means of encoding the
classical dynamics.
In this Section, we limit ourselves to transformations obtained by
normalisation of the $Y$ tensor corresponding to a long root,
since all interesting cases are obtained this way
(see Section \ExamplesSection). We thus use
$$
\sigma Y=-\eta_{\alpha\beta}T^\alpha\otimes T^\beta
+(\lambda,\lambda)-1+\sigma\;.\Eqn\SigmaY
$$

The fields in the coset $G/K\times\RR^+$
are parametrised by a generalised metric $G_{MN}$, which is at
the same time a group element in $G\times\RR^+$ and a symmetric
matrix.  The inverse metric will
be denoted $G^{MN}$.  Let
$G_{MN}$ transform as a tensor density with a weight $-2w$, which does
not necessarily equal $-2((\lambda,\lambda)-1)$, the canonical weight
of a tensor with two lower indices. We are looking for a density
${\cal L}$, containing two derivatives, that is invariant under
generalised diffeomorphisms, up to total derivatives. The weight $w$
will be determined. The result should be compared to known cases.

The introduction of a metric implies a preferred involution on
$\fg$, \ie, a local choice of embedding of the maximal compact
subalgebra ${\frak k}$.  Acting on generators in $R(\lambda)$, it is
$$
T^\alpha\mapsto -T^{\star\alpha}=-(GT^\alpha G^{-1})^t\;.\eqn
$$
The local generators in ${\frak k}$ and $\fg\ominus{\frak k}$ are
$T-T^\star$ and $T+T^\star$, respectively.
Let
$$
(G^{-1}\*_MG)^N{}_P=\Pi_{M\alpha}T^{\alpha
N}{}_P+\Pi_M\delta^N_P\;.\eqn
$$
When checking the transformations under generalised diffeomorphisms,
it is enough to check the inhomogeneous part, $\Delta_\xi
X\equiv\delta_\xi X-\LL_\xi X$. Using the form of the generalised
diffeomorphism above (with the appropriate weight), we obtain
$$
\eqalign{
\Delta_\xi\Pi_M&=-2w\*_M\*_N\xi^N\;,\cr
\Delta_\xi\Pi_{\alpha M}
&=\eta_{\alpha\beta}(T^\beta+T^{\star\beta})^N{}_P\*_M\*_N\xi^P\;.\cr
}\eqn
$$

It is often convenient to use the fact that the Killing metric is
invariant under the involution, so that
$\eta_{\alpha\beta}T^\alpha\otimes T^\beta
=\eta_{\alpha\beta}T^{\star\alpha}\otimes T^{\star\beta}$, and that
the adjoint index in $\Pi_{M\alpha}$ takes values in $\fg\ominus{\frak
k}$, \ie,
$$
\Pi_{M\alpha}T^\alpha=\Pi_{M\alpha}T^{\star\alpha}\;.\Eqn\PiSymmetric
$$
This immediately implies that the invariant tensors $Y$ and $Z$ fulfil
$$
G_{MM'}G_{NN'}Z^{M'N'}{}_{P'Q'}G^{P'P}G^{Q'Q}=Z^{PQ}{}_{MN}\;,\eqn
$$
and the corresponding identity for $Y$.
A calculation then shows that the combination
$$
\LL_0=\fr2A-B-2C
-{(\lambda,\lambda)\over(\lambda,\lambda)-\fr2}D\;,\Eqn\ABCDEq
$$
is necessary for the cancellation of various terms, where
$$
\eqalign{
A&=G^{MN}\eta^{\alpha\beta}\Pi_{\alpha M}\Pi_{\beta N}\;,\cr
B&=G^{PQ}T^{\alpha M}{}_PT^{\beta N}{}_Q\Pi_{\alpha N}\Pi_{\beta
M}\;,\cr C&=(T^\alpha G^{-1})^{MN}\Pi_{\alpha M}\Pi_{ N}\;,\cr
D&=G^{MN}\Pi_{ M}\Pi_{ N}\;.  }\eqn
$$
Note that the terms $A$ and $B$, which are the ones not containing the
scale variation $\Pi_M$, have a universal relative coefficient. In
previous formulations, based on traces in the fundamental rather than
using the Killing metric, the relative coefficient has been determined
on a case by case basis.

The remaining inhomogeneous transformation, modulo total derivatives,
is
$$
\Delta_\xi\LL_0=-4S^{\alpha
MN}{}_{PQ}G^{PS}\Pi_{S\alpha}\*_M\*_N\xi^Q\;,
\Eqn\RemainderEq
$$
where $S$ is the tensor of eq. (\SigmaEq).  This shows that $\LL_0$
gives the complete dynamics in all cases where ancillary
transformations are absent.

By cancellations of
inhomogeneous transformations,
the weight $w$ is also determined to be $w=(\lambda,\lambda)-\fr2$.
Note that this implies that the total weight of
$\LL$ is
$$
-2((\lambda,\lambda)-1)+2((\lambda,\lambda)-\fr2)=1\;, \eqn
$$
where
the first term comes from the two derivatives and the second one from
an inverse metric. This is the correct weight for partial integration,
in the sense that a divergence formed with a naked derivative is
covariant. Namely, consider a vector $V^M$ with weight $w^+$, Using
the section constraint, it is straightforward to show that
$$
\eqalign{
\*_M\LL_\xi V^M&=\xi^N\*_N\*_MV^M+(w^+-(\lambda,\lambda)+1)\*_N\xi^N\*_MV^M\cr
&\quad+(w^+-(\lambda,\lambda))\*_M\*_N\xi^NV^M\;.\cr}\eqn
$$
If $w^+=(\lambda,\lambda)$, the $\*^2\xi$ term vanishes, and this
equals $\LL_\xi\*_MV^M$. The divergence $\*_MV^M$ then is a scalar
density of weight $w^+-(\lambda,\lambda)+1=1$.

We do not know if it is possible in general to add terms to the
Lagrangian in order to cancel the remainder in eq. (\RemainderEq). It
was done for $E_8$ in ref. [\HohmSamtlebenIII]. That construction can
be extended to the class of models where $R(\lambda)$ is the adjoint
of a finite-dimensional Lie algebra (finite dimension is needed for the
adjoint to be a highest weight representation). Such extended
geometries are relevant in connection with compactification to 3
dimensions, and provide geometrisations of extensions of Ehlers
symmetry.

In the adjoint cases, using the generic form of the $Y$ tensor with
$T^{\alpha\beta}{}_\gamma=-f^{\alpha\beta}{}_\gamma$, the symmetry (\PiSymmetric) of
$\Pi$ can be expressed as follows. Let
$G_{\alpha\beta}=\phi\tilde G_{\alpha\beta}$, where $\tilde
G^{-1}d\tilde G$ takes values in $\fg$ (\ie, $\det\tilde G=1$), and
$\phi$ is a scalar density of the same weight as $G$ (\ie, $\phi=(\det
G)^{1/\dim\fg}$). Then,
$$
\tilde G^{\alpha\gamma}\Pi_{\gamma\beta}=\eta^{\alpha\gamma}\Pi_{\gamma\beta}\;.\eqn
$$
The remainder term of eq. (\RemainderEq) can in these cases be simplified using the
section constraint of the form
$$
\eta_{\kappa\lambda}f^{\kappa\gamma}{}_\alpha
f^{\lambda\delta}{}_\beta \*_{(\gamma}\*_{\delta)}=2\*_{(\alpha}\*_{\beta)}\;.\eqn
$$
Using the Jacobi
identity in the operator $S^\alpha$ then gives
$$
S_\alpha{}^{\delta\epsilon}{}_{\beta\gamma}\*_{(\delta}\*_{\epsilon)}
=f^\delta{}_{\beta\gamma}\*_{(\alpha}\*_{\delta)}\;,\eqn
$$
where the two derivatives act on the same (suppressed) parameter.
The remainder term is
$$
\Delta_\xi\LL_0=-4f^\beta{}_{\gamma\delta}G^{\gamma\epsilon}
\Pi_\epsilon{}^\alpha\*_\alpha\*_\beta\xi^\delta\;,\Eqn\AdjointRemainder
$$
where, as earlier, indices (except on $G$) are raised and lowered with
$\eta$, and only the presence of $G$ or its inverse are indicated
explicitly. It is cancelled by the variation of
$$
\LL_1=G^{\alpha\beta}\eta^{\gamma\delta}\Pi_{\alpha\gamma}\Pi_{\delta\beta}\;.\eqn
$$
This gives a complete (local) description of the dynamics in these
cases.

The characterisation of the ancillary transformations in cases where
$\fg$ is an affine algebra [\ENinePaper] also relied on a specific
rewriting of them, in those cases using the coset Virasoro generator
$L^{\hbox{\sixrm coset}}_{-1}$. If some $\LL_1$ is to be formed, it
seems likely that it will rely on that rewriting.  We have no further
insight in how to obtain an invariant Lagrangian in other
infinite-dimensional, \eg\ hyperbolic, algebras.

A comment on unimodular versus non-unimodular generalised metrics: In
\eg\ refs [\HohmSamtlebenI
--\HohmSamtlebenIII],
unimodular generalised metrics
are used. The density is provided by the ``external'' metric, and any
invariant expression will contain derivatives of (at least the
determinant of) the external metric. This scale can, if one wants, be
absorbed in the definition of a non-unimodular metric as above. In the
present context, we prefer to include the scale in the generalised
metric, since we are in a general situation where we do not want to
commit to a specific number of ``external'' coordinates. This applies
as long as the generalised metric, as defined here, carries a
non-trivial weight, \ie, as long as $(\lambda,\lambda)\neq\fr2$.

\section\ExamplesSection{Examples}We will take the opportunity to give
some examples, some connecting to known models, and some illustrating
how our formalism goes beyond already investigated cases.

The first example is the $E$ series, with $\lambda=\Lambda_1$. The
series contains the well-known exceptional geometries up to $E_8$, and
continues with $E_9$, where generalised diffeomorphisms have been
constructed [\ENinePaper] but the dynamics remains to be given. The
cases of $E_{10}$ and $E_{11}$ are of special interest.

\Figure\EnDynkinFigure{\epsffile{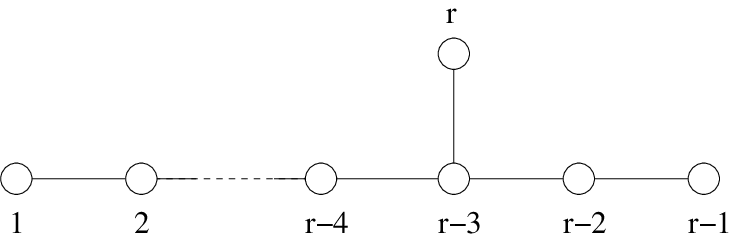}}{The Dynkin diagram for $E_r$.}

The highest irreducible module in $\wedge^2R(\Lambda_1)$ is
$R(2\Lambda_1-\alpha_1)=R(\Lambda_2)$.  It is obvious that no other
highest weight module occurs at the same height.  The next to
highest one appears (for $r\geq7$) at highest weight
$2\Lambda_1-\beta$, where $\beta=\sum_{i=1}^nb_i\alpha_i$,
$b_i=(2,\ldots,2,3,4,5,6,4,2,3)$ is the lowest root at level 2 with respect to node 1 (for
$r\geq8$).  The state
$$
\ketket{2\Lambda_1-\beta}
=\ket{\Lambda_1}\wedge e_{-\alpha_1}e_{-\beta+\alpha_1}\ket{\Lambda_1}
-e_{-\alpha_1}\ket{\Lambda_1}\wedge e_{-\beta+\alpha_1}\ket{\Lambda_1}
\eqn
$$
in $\wedge^2R(\Lambda_1)$ consists of two terms which are both
annihilated by all $e_i$, $i=2,\ldots,r$. The relative coefficient
assures that also $e_1\ketket{2\Lambda_1-\beta}=0$. This is the
highest weight state in the next-to-highest module
$R(2\Lambda_1-\beta)\in\wedge^2R(\Lambda_1)$.  (This may be
refined by stating the actual multiplicities of $2\Lambda_1-\beta$ in
$\wedge^2R(\Lambda_1)$ and in $R(\Lambda_2)$, and show that they
differ by 1, but it is not necessary for demonstrating that
$R(2\Lambda_1-\beta)\subset\wedge^2R(\Lambda_1)$.)

This means that ancillary transformations begin at degree $-3$ in the
M-theory grading, and at degree $-4$ in the type \II B grading, since
$(\beta,\Lambda_r)=3$ and $(\beta,\Lambda_{r-2})=4$. This illustrates
how the local transformations close for \eg\ $E_{10}$, although we
lack an action.

As a second example, take $\fg=A_r$, $\lambda=\Lambda_p$.  These cases
correspond to dimensional reduction of gravity.  The case
$\lambda=\Lambda_2$ was described in
refs. [\ParkSuhN,\StricklandConstable].  We only need to consider
$p\leq[{r+1\over2}]$, higher $p$ are related to lower by an outer
automorphism. There are (generically) two possible sections, one
$(p+1)$-dimensional ($p\geq2$) and one $(r-p+2)$-dimensional, arising
from following a gravity line to the left and to the right.  If we
decide going to the right (the $(r-p+2)$-dimensional section), all
cases are covered by also including higher $p$. There is an
R-symmetry $A_{p-2}$ when $p\geq3$.

\TextFigure\AnDynkinFigure{\epsffile{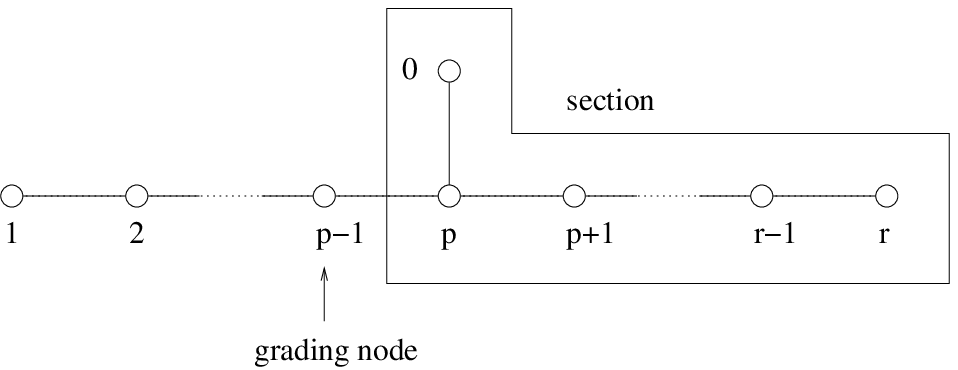}}{The Dynkin diagram
of $\scr A$ for $\fg=A_r$, $\lambda=\Lambda_p$,
with the subdiagram corresponding to one of the two sections.}

The field content is obtained by inspecting the grading of the adjoint
with respect to node $p-1$, \ie, its branching under $A_{p-2}\oplus
A_{r-p+1}\subset A_r$.  In addition to gravity, the fields are scalars
in the R-symmetry coset $SL(p-1)/SO(p-1)$ and $(p-1)$ 1-form potentials.
The Lagrangians\foot{We note that the case $r=3$, $p=2$ was excluded
in the analysis of ref. [\ParkSuhN]. For these values, one has
$(\lambda,\lambda)=1$, implying that the canonical weight of a vector
is $0$. Therefore, if one starts from a generalised metric
transforming as a tensor, it is not possible to form a density with
non-zero weight for integration. We
do not experience this problem, since we start from a generalised
metric which carries a non-tensorial weight.}
are given by eq. (\ABCDEq).

Another example where ancillary transformations are absent is
$\fg=C_r$, $\lambda=\Lambda_r$. The section is 2-dimensional. The
R-symmetry, obtained by deleting node $r-1$ and including the affine
node, is $C_{r-1}$. The field content in the coset is, apart from
2-dimensional gravity, scalars in the coset $Sp(2(r-1))/SU(r-1)$ and
$2(r-1)$ 1-form potentials in the fundamental of $Sp(2(r-1))$.

A word on solutions to the section constraint obtained as a gravity
line of short roots. Among such cases, there is no situation where
ancillary transformations are absent. The simplest example is when
$\fg=C_r$ and
$\lambda=\Lambda_1$, so that $R(\lambda)$ is the fundamental
representation. There is an $r$-dimensional section. This section is
at degree 0 in the grading with respect to node $r$, which is the
decomposition into $GL(r)$ modules. Since $c_r=1$, this is a
3-grading. Ancillary transformations will appear already at degree
$-1$, and remove everything except gravity from the coset
$Sp(2r)/SU(r)$.
This seems uninteresting.

The cases where the coordinate module is the adjoint
of a finite-dimensional Lie algebra are the only ones
where ancillary transformations are present and a Lagrangian is known.
Given a choice of section corresponding to a gravity line, both the
gauge parameters and the fields are deduced from the corresponding
grading. This grading is defined by the node or nodes outside of, but
connected to, the gravity line. As in Section~\SectionDiffSection, call this node (or one of them)
number $j$.
The relevant next-to-highest module in $\wedge^2 R(\lambda)$,
with $\lambda=\theta$, is $R(\theta)$ itself. Ancillary
transformations appear at degree $-c_j$, where $c_j$ is the degree of
$\theta$ in the grading, \ie, the Coxeter label of the grading node
$j$. This is also the lowest degree appearing in the adjoint
representation, which implies that ancillary transformations commute.
The ancillary transformations are then always a 1-form (with respect
to the section), which is equivalent to a section-constrained adjoint
element. The corresponding element in a (linearised) coset can always
be shifted away by an ancillary transformation. The parameters of
generalised diffeomorphisms at degree $0$ and lower become irrelevant;
such transformations can always be absorbed in an ancillary
transformation. The pattern from $E_8$ exceptional geometry
[\HohmSamtlebenIII,\CederwallRosabal] is
repeated. The precise content of ``matter'' fields depends on
details.

For example, the $E_7$ theory
with the $7$-dimensional section (the same gravity
line as ordinary exceptional $E_7$ geometry, but with the coordinate
module in the opposite end) contains a $4$-form gauge
potential in addition to gravity. Since the grading of the adjoint
with respect to the exceptional node is a 5-grading, there are no
other physical fields.

\section\ConclusionsSection{Conclusions}We have presented a unified
formalism for dealing with extended geometry, only relying on the
choice of structure group and coordinate module. The treatment is so
far local.

Inclusion of an ``external'' space with ordinary diffeomorphisms has
intentionally been left out, since the actual field content depends on
other input, \eg\ supersymmetry or a knowledge of the duality group
present in a certain compactification.
Other fields than the ones described here (for dimension of the
external space $D\geq4$) would then be introduced through a tensor
hierarchy in cases where ancillary transformations are absent. The
techniques are straightforward.

The generalisation of the formalism to include supergeometries
[\CederwallDoubleSuperGeometry] should
be straightforward. The Dynkin diagram is replaced by the Dynkin
diagram of a superalgebra, which plays the r\^ole of structure algebra
for the generalised superdiffeomorphisms. A super-section constraint
restrict the super-section to a ``supergravity line'' corresponding to
the structure superalgebra ${\frak gl}(d|s)$ of ordinary
supergeometry.
However, to be concrete about field content and symmetries,
information from the representation theory of Kac--Moody superalgebras
will be needed, and not much seems to be known even concerning the
(infinite-dimensional) super-extensions of $E_r$. Neither is it
clear how dynamics should be formulated.

Many questions remain to be investigated. Among the most pressing ones
is to determine the dynamics, if not in all cases, at least in some
important ones, such as affine and hyperbolic cases. It is also
desirable to obtain a better understanding of finite transformations,
which in double field theory are reasonably well understood
[\HohmZwiebachLarge,\BermanCederwallPerry--
\CederwallDuality],
but which have been elusive in the exceptional cases, for fundamental
or technical reasons. A generic treatment is desirable.

In section \DynamicsSection, we noted that there seems to be a
singularity for $(\lambda,\lambda)=\fr2$. In general, this is not
serious, but only a sign that the generalised metric (defined as a
density, as in section \DynamicsSection) carries no weight. If some
external space is present, a density may be introduced that
compensates, and (part of) an action may be written.
A peculiar observation is that this happens precisely for the case
$E_{11}$, where no external space should be present. The
interpretation is not clear to us, but may point towards a difficulty
with formulating an action in that case.

The treatment in the present paper has its focus on transformations
and their closure, together with invariant dynamics in some
cases. Closure of the algebra of generalised diffeomorphisms
(\ie, absence of ancillary transformations) ensures
covariance 
[\CederwallRosabal], but when ancillary transformations are present
tensors cannot be defined unless the transformations are made
field-dependent. This was done for $E_8$ in ref. [\CederwallRosabal],
and can easily be extended to all the finite-dimensional adjoint cases
with the methods of the present paper. We do not know
if this is possible in general. Without a concept of covariance, it is
not possible to define truly geometric entities like torsion,
curvature etc., which would be desirable. Again, the most urgent cases
are $E_r$, $r\geq9$.

Our formalism is based on the weight space decomposition of modules of
the Lie algebra $\fg$, which provides a real basis when $\fg$ is of
split real form. This is the real form typically used in double and
exceptional field theory. We have not tried to analyse in detail what
happens for other choices of real form, like the orthogonal groups in
ref. [\HohmMusaevSamtleben]. For a real representation 
there should be no problems with the definition of the generalised
diffeomorphisms and the section constraint, which are built using real
invariant tensors. The solution of the section constraint in terms of
a real gravity line seems to demand that the real form is obtained
from the split one through redefinitions of generators not affecting
the ${\frak gl}(d)$ subalgebra. The dynamics must be completely
reconsidered when the compact subgroup is changed. 

The fermionic extensions of the Kac--Moody algebra $\fg$ used in the present work contain
$\fg$ itself at level $0$, the coordinate module $R(\lambda)=R_1$ at level $-1$, and the section constraint
module $R_2$ at level $-2$. They are contragredient (in particular, Borcherds superalgebras),
which means that the modules at the positive levels are dual to those at the corresponding negative levels.
Alternative extensions,
essentially agreeing at positive levels, but with different (larger)
modules at negative levels, are the tensor hierarchy algebras
[\PalmkvistTensor,\CarboneCederwallPalmkvist]. These algebras seem to
have a deeper connection to extended geometry.
A peculiar aspect of the tensor hierarchy algebras corresponding to
infinite-dimensional extended geometries is
the appearance of certain additional elements at level $0$, \ie,
together with the algebra $\fg$. They have already been demonstrated
to be important [\BeyondEEleven,\ENinePaper]. It is quite possible that such
transformations should be considered part of the structure algebra,
when generalised diffeomorphisms are constructed, and there might be a
subtle connection between these elements and the ancillary transformations.
Issues like these have been ignored in the present work, but seem to
point towards interesting potential development. 

\acknowledgements We would like to thank Guillaume Bossard, Axel
Kleinschmidt and Henning Samtleben for discussions, and for the
collaboration [\ENinePaper] that initiated this work. The research is
supported by the Swedish Research Council, project no. 2015-04268.


\refout

\end

\line{\hrulefill}

An even shorter way of deriving the form of the $Y$ tensor is to ask
that the symmetric part gives the minimal orbit, and that the
antisymmetric part annihilates the highest antisymmetric module. It
has highest weight $2\lambda-\alpha$, where $\alpha$ is the simple
root dual to the fundamental weight $\lambda$, and the highest state
is
$$
\ket{\ket{2\lambda-\alpha}}
=\ket\lambda\otimes\ket{\lambda-\alpha}-\ket{\lambda-\alpha}\otimes\ket\lambda
\;.\eqn
$$
Direct calculation gives $C_2(R(2\lambda-\alpha))=C_2(R(2\lambda))-2$,
which dictates the term $-1+\sigma$.

Also, similar considerations show that the antisymmetric part of $Y$
vanishes if and only if the antisymmetric product $\wedge^2R(\lambda)$
is irreducible.

A strict proof demands that one shows that
$$
\mu\prec\nu\Rightarrow C_2(\mu)\leq C_2(\nu)-2\;.\eqn
$$
This is done by direct calculation. Setting $\mu=\nu-\alpha$,
$\alpha>0$,
gives
$$
C_2(\nu-\alpha)=C_2(\nu)-(\alpha,\varrho)-(\alpha,\nu)
+\fr2(\alpha,\alpha)\;.\eqn
$$
We have $\fr2(\alpha,\alpha)\leq1$ and
$-(\alpha,\varrho)\leq-1$. $(\alpha,\nu)$ is generically $\geq0$,
but if it is smaller than 2, $\nu-\alpha$ can not be a dominant
weight. This is obvious for $(\alpha,\nu)=0$, and for $(\alpha,\nu)=1$
it follows from $\nu-\alpha$ then being the Weyl reflection of $\nu$
in the hyperplane orthogonal to $\alpha$.